\documentclass[letterpaper]{article} % DO NOT CHANGE THIS
\usepackage{aaai2026}  % DO NOT CHANGE THIS
\usepackage{times}  % DO NOT CHANGE THIS
\usepackage{helvet}  % DO NOT CHANGE THIS
\usepackage{courier}  % DO NOT CHANGE THIS
\usepackage[hyphens]{url}  % DO NOT CHANGE THIS
\usepackage{graphicx} % DO NOT CHANGE THIS
\usepackage{natbib}  % DO NOT CHANGE THIS AND DO NOT ADD ANY OPTIONS TO IT
\usepackage{caption} % DO NOT CHANGE THIS AND DO NOT ADD ANY OPTIONS TO IT
\usepackage{algorithm}
\usepackage{algorithmic}

\usepackage{subcaption}
\usepackage{booktabs}
\usepackage{amsmath}
\usepackage{multirow}
\usepackage{xcolor}
\usepackage{etoolbox}
\usepackage{enumitem}
\usepackage{etoolbox}
\usepackage{environ}
\usepackage{xparse}
\usepackage{ragged2e} 
\usepackage[bb=boondox,bbscaled=.95,cal=boondoxo]{mathalfa} 
\newcommand{\blockquoteauthor}{} % temporary storage
\newenvironment{blockquote}[1][]%
{%
  \renewcommand{\blockquoteauthor}{#1}%
  \begin{list}{}{\leftmargin=1em \rightmargin=1em}
  \item\relax
  \small
  \setlength{\parindent}{1em}
  \justifying
}%
{%
  \end{list}
  \ifstrempty{\blockquoteauthor}{}{%
    \vspace{-0.5em}
    \begin{flushright}
      \small\itshape-- \blockquoteauthor
    \end{flushright}
  }%
  \smallskip
}

\usepackage{newfloat}
\usepackage{listings}
\DeclareCaptionStyle{ruled}{labelfont=normalfont,labelsep=colon,strut=off} % DO NOT CHANGE THIS
\floatstyle{ruled}
\newfloat{listing}{tb}{lst}{}
\floatname{listing}{Listing}
\nocopyright % -- Your paper will not be published if you use this command
\title{Rabble-Rousers in the New King's Court: Algorithmic Effects on Account Visibility in Pre-X Twitter}
\author {
    Alexandros Efstratiou\textsuperscript{\rm 1},
    Kayla Duskin\textsuperscript{\rm 1},
    Kate Starbird\textsuperscript{\rm 1},
    Emma S. Spiro\textsuperscript{\rm 1}
}
\affiliations {
    \textsuperscript{\rm 1}University of Washington\\
    \{aefstra, kduskin, kstarbi, espiro\}@uw.edu
}
\begin{document}

\maketitle

\begin{abstract}

Algorithmic effects on social media platforms have come under recent scrutiny, with several studies reporting that right-leaning accounts tend to receive more exposure.
In this paper, we expand upon this body of work using data collected from user feeds after Twitter's change of ownership but before its re-branding to X.
We replicate findings from prior work regarding the increased exposure of right-leaning accounts to wider audiences in algorithmically curated compared to reverse-chronological feeds, and, crucially, we further unpack this effect to illuminate what correlated (and did not correlate) with these differences.
Our results reveal that right-leaning accounts benefited not necessarily due to their political affiliation, but likely because they behaved in ways associated with algorithmic rewards; namely, posting more agitating content and receiving attention from the platform's owner, Elon Musk, who was the most central network account.
We also demonstrate that legacy-verified accounts, like businesses and government officials, received less exposure in the algorithmic feed compared to non-verified or Twitter Blue-verified accounts.
We discuss implications of these findings for the intersection between behavioral incentives for algorithmic reach and the health of online discourse. 

\end{abstract}


\section{Introduction}

\begin{blockquote}[Elon Musk, soon after his Twitter acquisition\footnote{\url{https://archive.ph/PEG2d}}]
    ``New Twitter policy is freedom of speech, but not freedom of reach.''
\end{blockquote}

The role of algorithms in amplifying divisive and problematic content is a question of both societal~\cite{pariser_filter_2011} and academic concern~\cite{ribeiro_auditing_2020}.
Recently, debates have sparked around whether recommendation algorithms disproportionately amplify or suppress content from specific political camps.
For example, the Federal Trade Commission under the second Trump administration launched an investigation into alleged ``censorship'' of conservative voices,\footnote{\url{https://archive.ph/xcShX}} while other reports suggest right-leaning accounts actually receive outsized amplification~\cite{graham_computational_2024}.
Yet, others, including former Twitter employees~\cite{messing_what_2023}, have argued that algorithmic effects are deceptively difficult to measure because of algorithms' inherent reliance on user preferences that shape what these algorithms learn~\cite{ribeiro_amplification_2023}.

The challenge and urgency of this debate has prompted research aimed at isolating and characterizing algorithmic effects in political contexts. Much of this work has relied on the use of automated accounts, or ``bots'' to capture algorithmically recommended content \cite{ye_auditing_2025, Duskin2026-new, Bandy2021-lg}.
Recent work has also incorporated ``counterfactual bots'' to control for baseline user behavior~\cite{hosseinmardi_causally_2024}.
%In line with these arguments, recent work has focused on isolating algorithmic effects by deploying and observing ``counterfactual bots'' while controlling for baseline user behavior~\cite{hosseinmardi_causally_2024}.
That is, a pair of bots, one of which is instructed to behave like a real user and the other instructed to randomly follow algorithmic recommendations, can be compared to estimate content amplified beyond baseline user preferences.
% the kinds of content that an algorithm may amplify above and beyond what a user would organically opt for.
% However, another approach we can take is the idea of ``counterfactual feeds''; that is, instead of comparing the behaviors of different simulated users while controlling for organic behavior, we can compare empirical differences between different feed configurations for the same organic users.
% This approach cannot reliably offer insights on algorithmic amplification, as this continues to depend on behaviors that may be learned by other users.
% However, it is a good measurement of algorithmic priors, which offer a window into how global interactions of algorithmic parameters and user behaviors affect what is visible for other users of a platform.

In this work, we take an approach of ``counterfactual feeds''; that is, taking the same user at the same time, what do their algorithmic feeds look like compared to if we ``switched off'' the algorithm?
Notably, our data is entirely from real Twitter users, allowing us to compare the algorithmic and chronological feeds within a wholly realistic setting.
Additionally, while prior work has made significant progress in observing \textit{how} content is disproportionately amplified, our study dives into the question of \textit{why} this may be occurring.
That is, although surface-level differences between the visibility of left- or right-leaning accounts may indeed exist, we also question whether these differences may be driven by fundamental differences in how these accounts behave. 

Given the recent academic and political interest in the potential of algorithmic effects to drive political bias, we focus on political account visibility using a dataset that collects posts from real users' (as opposed to automated audit accounts') Twitter feeds.
These data were collected immediately prior to the platform's re-branding to $\mathbb{X}$, at a time of significant change within the organization.
We not only characterize differences in exposure but also attempt to disentangle the behaviors that could be driving discrepancies leading to conclusions of political bias.

% \paragraph{Research questions.}
Specifically, we pose the following research questions:

\begin{enumerate}[align=left]
    \item[\textbf{RQ1.}] How does algorithmic ranking impact the visibility of political accounts?
    \item[\textbf{RQ2.}] What are the account characteristics of algorithmic beneficiaries and losers?
    \item[\textbf{RQ3.}] Which underlying differences of these characteristics between political accounts explain visibility differences?
\end{enumerate}

% \paragraph{Methods overview.}
% We use data from~\citet{milli_engagement_2025}'s study on algorithmic elevation of emotional and outgroup-derogating content, who collected simultaneous algorithmic and chronological Twitter feeds from participants recruited from an online panel.
% Building upon their work, we focus our analyses on account level amplification, rather than the tweet level.
% Crucially, we find that this account-level context --- such as an author's social proximity to the network center --- may have played an important role in content recommendations.
% % This allows us study their proximity to the network center that, as we argue throughout the paper, may play an important role in content recommendations.
% We derive several features for these accounts, such as the degree to which they posted political or agitating tweets and their relationship to the most central node in the network in terms of eigenvector centrality.
% We then determine the extent to which these features were associated with gains or losses in exposure to new users between the reverse-chronological and algorithmic feed.

\paragraph{Main findings.}

We use data from~\citet{milli_engagement_2025}'s study on algorithmic elevation of emotional and outgroup-derogating content, who collected simultaneous algorithmic and chronological Twitter feeds from participants recruited from an online panel.
Building upon their work, we focus our analyses on account level amplification, rather than the tweet level, so that we can also capture account characteristics like social proximity to the network center.

We find that, during the data collection window (in February 2023), right-leaning accounts enjoyed greater increase in visibility in participants' algorithmic feeds compared to their chronological feeds than left-leaning and neutral accounts.
This held irrespective of whether the feed belonged to a self-identified Democrat, Republican, or Independent participant.
The algorithmic feed showed substantially higher centralization of influence, with much of this driven by the platform owner, Elon Musk, receiving disproportionately more exposure in the algorithmic compared to the chronological feed.
Subsequently, gains in algorithmic visibility were higher for accounts that Elon Musk replied to or retweeted, and accounts that posted more agitating content.
Losses in visibility were associated with posting more political content, being legacy-verified, and leaning left politically. Twitter Blue verification did not change visibility compared to unverified accounts.
Importantly, when controlling for attention from Elon Musk, verification status, and posting styles, the gains in visibility observed in right-leaning accounts disappeared.

% \begin{itemize}
%     \item During the window of data collection (in February 2023), right-leaning accounts enjoyed greater increase in visibility in the algorithmic feed compared to the chronological feed than left-leaning and neutral accounts.
%     \item Further analyses suggest that this held irrespective of whether the feed belonged to a self-identified Democrat, Republican, or Independent.
%     \item The platform's owner, Elon Musk, received disproportionately more exposure in the algorithmic feed compared to the chronological feed.
%     \item Gains in visibility in the algorithmic feed were higher for accounts that posted more agitating content and accounts that Elon Musk replied to or retweeted.
%     \item Losses in visibility were associated with posting more political content, being legacy-verified, and leaning left politically. Twitter Blue verification did not change visibility compared to unverified accounts.
%     \item When controlling for attention from Elon Musk, verification status, and posting styles, the gains in visibility observed in right-leaning accounts disappeared.
%     \item Our analyses show that right-leaning accounts possibly enjoyed higher visibility \textit{because} they used more agitating language and received more attention from Elon Musk.
% \end{itemize}

\paragraph{Contributions.}
These findings challenge the notion that the Twitter algorithm necessarily amplifies right-leaning accounts due to their political stance; rather, the results are more consistent with the explanation that right-leaning accounts may post more agitating content or receive attention from the platform's owner --- both of which are linked to algorithmically increased exposure.
This has implications for perverse incentives, especially given subsequent changes that introduced monetization.
The increased prominence of problematic content and disproportionate centralization of algorithmic influence necessitate increased scrutiny on these exposure mechanisms, and raise doubts about Twitter/X's claimed goal to act as the ``digital town square.\footnote{\url{https://www.pbs.org/newshour/politics/musk-doesnt-want-twitter-free-for-all-hellscape-he-tells-advertisers}}''

\section{Related Work}

% Our work is a description of algorithmic outcomes around a period for which we have a better understanding of its inner workings, and so cannot speak to amplification above and beyond user preferences.
Our work is a \textit{description} of user experience under algorithmic contexts, in this case on Twitter, as opposed to a strict causal assessment of algorithmic effects above and beyond user preferences.
Notwithstanding, it is best situated within a growing body of work that has conducted algorithmic audits to understand the kinds of content and accounts that are most-often recommended on social media platforms.
% ~\cite{galeazzi_revealing_2025}.

\subsection{Sockpuppet and Automated Audits}

Among the influential recent work sparking interest in algorithmic audits is \citet{ribeiro_auditing_2020}, which traced video recommendations on YouTube to chart radicalization pathways from pseudo-intellectualism to the alt-right.
These types of studies proliferated on YouTube~\cite{haroon_auditing_2023, ibrahim_youtube_2023, hosseinmardi_causally_2024} primarily because its API used to offer endpoints that enabled this kind of study, something which is no longer the case.
Since then, multiple other studies have emulated similar methods by deploying automated accounts that simulate users (aka, sockpuppets) and observing the content recommended to them.
This includes studies on Twitter's algorithmic timeline, and geolocation-based SERP audits for COVID-19 misinformation on YouTube~\cite{jung_algorithmic_2025}, among others.
The benefit of these automated audits is that they are not obfuscated by user activity, allowing for the study of algorithmic baselines or priors.

On Twitter specifically, \citet{Bandy2021-lg} deployed ``archetype puppets" to emulate users from varying communities, and found that the platform's algorithmic feed increased exposure to niche partisan accounts while decreasing bipartisan sources. \citet{duskin_echo_2024_new} conducted an audit of the platform's friend recommender by deploying sockpuppets that grew their network with or without input from the `Who to Follow' recommender. They found that ``user preferences'', i.e., the stochastic expansion that ignored the recommender, resulted in more homogeneous networks than algorithmic recommendations.
In another sockpuppet study, \citet{Duskin2026-new} found that Twitter's algorithmic feed produced a small, but consistent skew toward right-leaning authors.
Another recent study by \citet{ye_auditing_2025} deployed 120 sockpuppets on $\mathbb{X}$ during the 2024 US Presidential Election, finding that right-leaning accounts benefited the most from out-of-network exposures.

\subsection{User-Based Audits}

Despite their benefits, a common critique of automated audits is that algorithms effectively reflect learned, aggregated user preferences; without controlling for user behavior, one cannot say that the algorithm amplifies specific kinds of content~\cite{lam_sociotechnical_2023,ribeiro_amplification_2023}.
To that end, several works have instead conducted user experiments to gauge what is seen by real users when the algorithm is ``turned off'' on platforms like Facebook and Instagram~\cite{guess_how_2023} or $\mathbb{X}$~\cite{wang_lower_2024}, while others have deployed sockpuppets modeled after real users and compared them to others behaving stochastically, for example on YouTube~\cite{hosseinmardi_causally_2024}.

One of the largest such studies was conducted by the Twitter team in collaboration with academics~\cite{huszar_algorithmic_2022}.
This work, which randomized $\sim$2M Twitter users into a reverse-chronological feed, found that algorithmic amplification favored right-leaning politician accounts and news sources compared to left-leaning ones, although it did not find evidence of amplification for users belonging to extreme groups.
Most similar to our work, \citet{milli_engagement_2025}, whose data we use in this study, collected both the engagement and reverse-chronological Twitter feeds of the same users at the same time.
They found that the algorithmic feed featured more emotional and outgroup-derogating content, although this was not necessarily content that users reported preferring. Here, we expand beyond this content-centered analysis to consider how user-level characteristics and interaction patterns are associated with visibility within the algorithmic feed.  

\subsection{Present Study}

Despite notable contributions and a growing body of prior work, we still lack an adequate understanding of the \textit{kinds} of accounts that algorithmic manifestations benefit.
Indeed, few works focus at the account level~\cite{ye_auditing_2025}. We argue for addressing this gap because 1) it allows us to capture network effects, specifically, the proximity of accounts to the most central network node, that are inherently built into algorithmic recommendations and 2) it allows us to better understand social media \textit{incentives} in gaining influence, which are paramount for regulators and legislators.
Moreover, although some works offer rich descriptions of what kind of content may benefit algorithmically~\cite{milli_engagement_2025}, we can paint a fuller picture with more research that considers previously unexamined dimensions.

\section{Methods}

Here, we briefly describe the dataset we use and how we further transform the data for our analyses.

\subsection{Dataset}
We use a dataset made available by~\citet{milli_engagement_2025}, who collected both the reverse-chronological and ``For You'' (henceforth, ``engagement'' as per the original paper) Twitter feeds of 806 US residents recruited on the CloudResearch Connect panel platform between February 11th-27th, 2023. Though the collection window was short and the resulting dataset is relatively small, these data offer insight into a particular salient moment in the platform's history: they were collected (1) as the platform was changing identity in response to new ownership and (2) one month before Twitter released code for its recommendation algorithm on March 31st 2023,\footnote{\url{https://github.com/twitter/the-algorithm}} making it highly likely that our observations are driven by the same or very similar version of that algorithm.
This allows for a unique juxtaposition between what the algorithm was \textit{built to do} versus what it \textit{did do}, namely, the stated purpose of recommending more interesting content versus the potential for amplifying more problematic content or over-centralizing recommendations among a few important accounts.
%which can allow for more direct comparisons.

Across both feeds and all participants, this dataset captures 205k potential exposures to 171k unique tweets authored or retweeted by 63.4k unique accounts.
We direct the reader to the original paper for detailed participant demographics.
However, given the politically-adjacent focus of this paper, we clarify that the dataset is heavily skewed with 76.7\% of participants identifying as left-leaning and 23.3\% as right-leaning, suggesting that the sample may not be nationally representative.\footnote{\citet{milli_engagement_2025} also report that the Twitter user population skewed Democrat in a 2020 ANES study of a nationally representative sample, but not to the degree of this dataset.}
Therefore, we make balancing adjustments where necessary.
The dataset does not contain any promotional or ad tweets.

\subsection{Network Structure}
We draw bipartite networks between a participant P and a Twitter account A if a tweet from A appears in P's feed, such that we form directed edges $P\rightarrow~A$.
Thus, unweighted networks reflect account exposures to unique participants, whereas weighted networks also consider the number of exposures of A to P.
We do not consider tweets if they are only shown to participants as a quoted or replied-to tweet, but we do consider replies or quote tweets themselves.

\begin{figure}[t!]
    \centering
    \begin{subfigure}[t]{0.49\linewidth}
        \centering
        \includegraphics[width=0.99\linewidth]{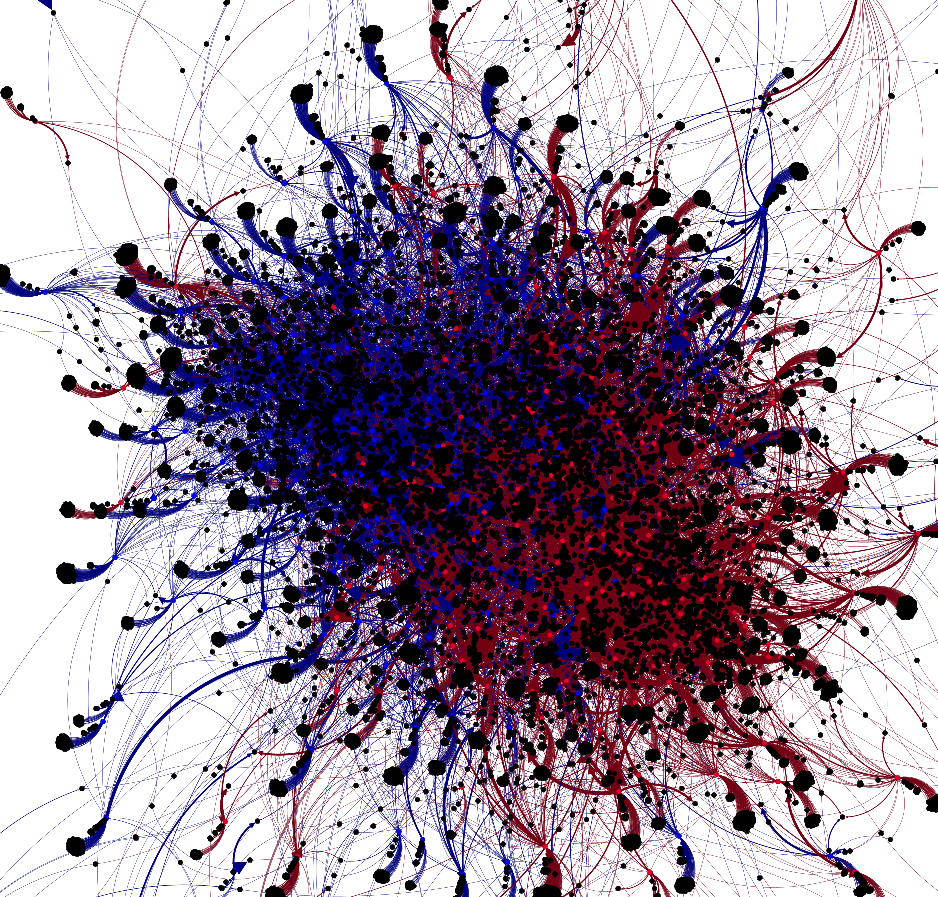}
        \caption{Reverse-chronological feed}
        \label{subfig:networks_chron}
    \end{subfigure}
    \begin{subfigure}[t]{0.49\linewidth}
        \centering
        \includegraphics[width=0.99\linewidth]{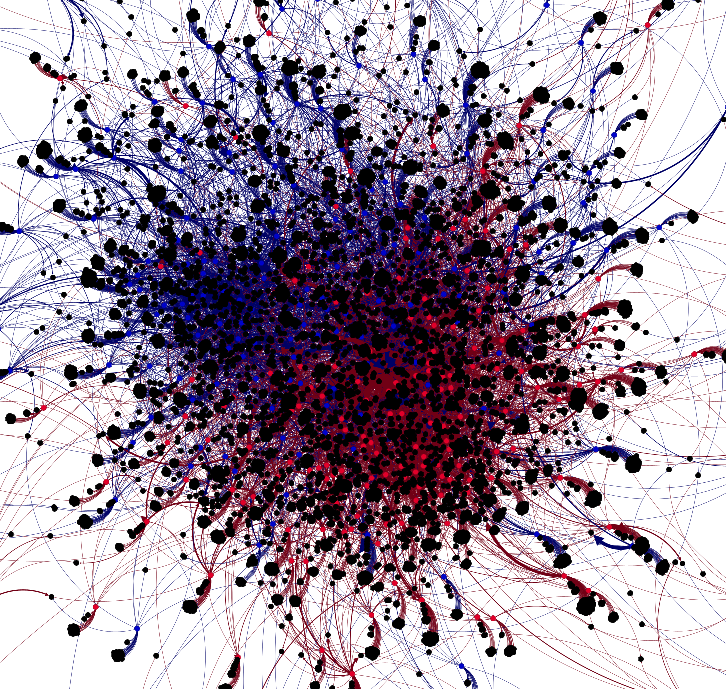}
        \caption{Engagement feed}
        \label{subfig:networks_eng}
    \end{subfigure}
    \caption{Balanced network visualizations.}
    \label{fig:networks}
\end{figure}

\paragraph{Participant matching.}
To address the political skew of the sample, we match the minority right-leaning participants to a subset of left-leaning ones, using the demographic variables that~\citet{milli_engagement_2025} obtained.\footnote{See \url{https://github.com/smilli/twitter/blob/main/DATA.md} for potential responses.}
We form participant vectors consisting of the categorical variables race, gender, and reason for using Twitter (e.g., entertainment, to stay informed, etc.), and the ordinal variables education level, age group, and annual household income.
Categorical variables are one-hot encoded, and all (resulting) variables are equally weighted.
We compute pairwise cosine distances between all right-leaning to left-leaning participants, and perform nearest-neighbor matching without replacement (such that each right-leaning user is matched to exactly one unique left-leaning user).
Whenever we refer to a ``balanced network'' henceforth, we mean a network based on this matched set of participants.
We show the (unweighted) balanced networks based on the reverse-chronological and engagement feeds in Figure~\ref{fig:networks}.
Results presented in the next section are largely consistent with analyses utilizing the full sample.
We demonstrate the success of our matching procedure in the Appendix.

\section{Feed Differences in Influential Nodes}

In this section, we provide an overview of the types of accounts that gained prominence when switching from the chronological to the engagement feed.

\subsection{Descriptive Statistics}

We begin with a description of the two unweighted feed configurations in Table~\ref{tab:net_stats}.
To compute partisan assortativity, we project the networks such that an edge forms between two participants if they are exposed to the same account.
To compute (in-)degree centralization, we implement~\citet{borgatti_network_1997}'s method for bipartite graphs that computes the theoretical maximum degree centralization by accounting for the cardinality of different vertex sets instead of relying on a unipartite star graph as follows:

\begin{equation}
    \frac{\sum[C(p^*) - C(p_i)]}{(n_i + n_0)n_i - 2(n_i + n_0 - 1)}
\end{equation}

Where $p\in V$, $V$ is the vertex set for mode-2 nodes (i.e., accounts), $p^*$ is the highest in-degree in $V$, $n_i$ is the cardinality of $V$, and $n_0$ is the cardinality of the set of mode-1 nodes (i.e., participants).

\begin{table}[t!]
    \centering
    \begin{tabular}{lrr}
    \toprule
        \textbf{Attribute} & \textbf{Chronological} & \textbf{Engagement} \\
        \midrule
        Assortativity & \textbf{0.15} & 0.06 \\
        Centralization & 0.24 & \textbf{0.46} \\
        \textit{N} mode-2 nodes & \textbf{22.9k} & 11.8k \\
        \textit{N} edges & \textbf{31.9k} & 17.7k \\
        \bottomrule
    \end{tabular}
    \caption{Network statistics. \textit{N} mode-1 nodes (participants) is 376 in both feeds (188 left- and 188 right-leaning).}
    \label{tab:net_stats}
\end{table}

Although there was a larger set of exposed accounts (and, by extension, $P\rightarrow A$ edges) in the reverse-chronological feed, this feed also showed higher partisan assortativity, meaning that there was more partisan homogeneity in the accounts that participants were exposed to.
This is in line with several recent works that suggest partisan sorting may mostly arise due to user preference rather than algorithmic recommendations~\cite{chouaki_what_2024,duskin_echo_2024_new,robertson_users_2023}.
However, we also notice much higher centralization in the engagement feed, suggesting that exposure was more concentrated among a few important accounts.
We explore this finding next.
The patterns we observe are identical when preserving the entire participant sample (\textit{N} = 806).

\subsection{Gains and Losses in Prominence}

To investigate which accounts gained and lost the most prominence when switching from the reverse-chronological to the engagement feed, we first classify their political leaning.
We assign a score $\lambda$ to each account based on the number of right-leaning participants that followed them divided by the total number of participants in the balanced network, such that 0 means an account was followed solely by left-leaning participants and 1 means it was followed only by right-leaning ones.
Where a participant followed or unfollowed an account during the observation period, we take the most recent status ($< $ 0.5\% of cases).

Since some accounts were followed by more participants than others and thus had lower classification error rates, we also derive binomial proportion (Wilson) confidence intervals at the 80\% confidence level for each account and classify them as follows:

\[
\text{leaning} =
\begin{cases}
\text{right} & \text{if } \lambda > 0.5 \text{ and } CI_{\text{lower}} > 0.5 \\
\text{left} & \text{if } \lambda < 0.5 \text{ and } CI_{\text{upper}} < 0.5 \\
\text{neutral} & \text{otherwise}
\end{cases}
\]

In other words, accounts for which confidence intervals span the 0.5 midpoint are classified as neutral.
We choose an 80\% CI as a reasonable trade-off between true positives and false negatives that does not over-classify while still allowing us to label $\sim$10\% of the accounts in the sample as left or right.
We also perform robustness checks at different confidence levels (70-95\% in 5\% increments)\footnote{We do not perform confidence analyses for the non-balanced graph, as left accounts need more samples to be confidently classified which artificially inflates the class in-degree.} 
%differing left-right thresholds resulting from sample imbalance adjustment artificially inflate left-classified in-degrees and deflate right-classified ones. 
as well as classifications with strict cut-offs instead of confidence intervals ($0.5<|>0.5$, $0.45<|>0.55$ and $0.4<|>0.6$ for left and right, respectively, and proportional cut-offs for the non-balanced graph) and find consistent results for the analyses that follow.
All three leaning classes followed similar daily activity patterns in terms of unique users and number of tweets posted (see Appendix).

In Figure~\ref{fig:ccdfs}, we plot the Complementary Cumulative Distribution Functions (CCDFs) for node in-degrees and eigenvector centralities across weighted and unweighted versions of the graphs.
For both metrics and across any configuration, we observe that left-leaning accounts trended more influential in reverse-chronological feeds (with the exception of the most influential right-leaning node, which corresponds to Elon Musk's account, overtaking left-leaning nodes in the unweighted configurations; Elon Musk was the most followed account by participants in the sample and second-most followed overall).
% The exception is overtaking of the most influential right-leaning node in the unweighted configurations; this corresponds to Elon Musk's account, which is the most followed account by participants in the sample and the second-most followed account overall.
However, this pattern was reversed in the engagement feed.
Right-leaning accounts received consistently more exposure (higher in-degrees) and influence (higher eigenvector centrality) in all engagement feed configurations. 
Looking at the x-axes specifically, we observe that right-leaning accounts tended to gain in-degrees and eigenvector centrality in both weighted and unweighted versions, while left-leaning accounts tended to lose both metrics in both versions.
Neutral (unclassified) accounts gained in unweighted versions but lost in weighted ones, suggesting that, while the engagement feed resulted in them being exposed to more users, their raw number of exposures was reduced.

\begin{figure*}[t!]
    \centering
    \begin{subfigure}[t]{0.49\textwidth}
        \centering
        \includegraphics[width=0.99\textwidth]{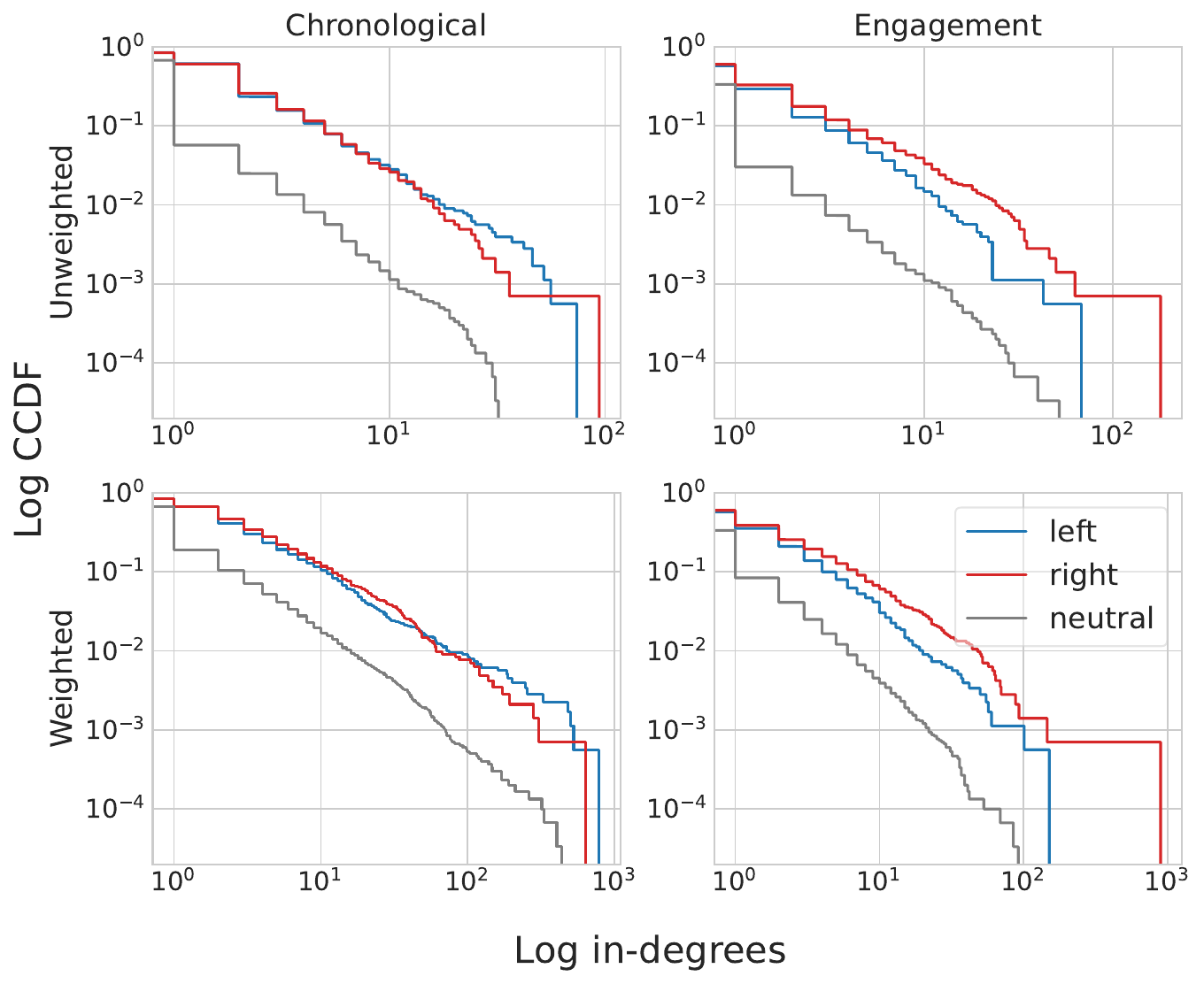}
        \caption{In-degree CCDFs.}
        \label{subfig:ccdfs_deg}
    \end{subfigure}
    \begin{subfigure}[t]{0.49\textwidth}
        \centering
        \includegraphics[width=0.99\textwidth]{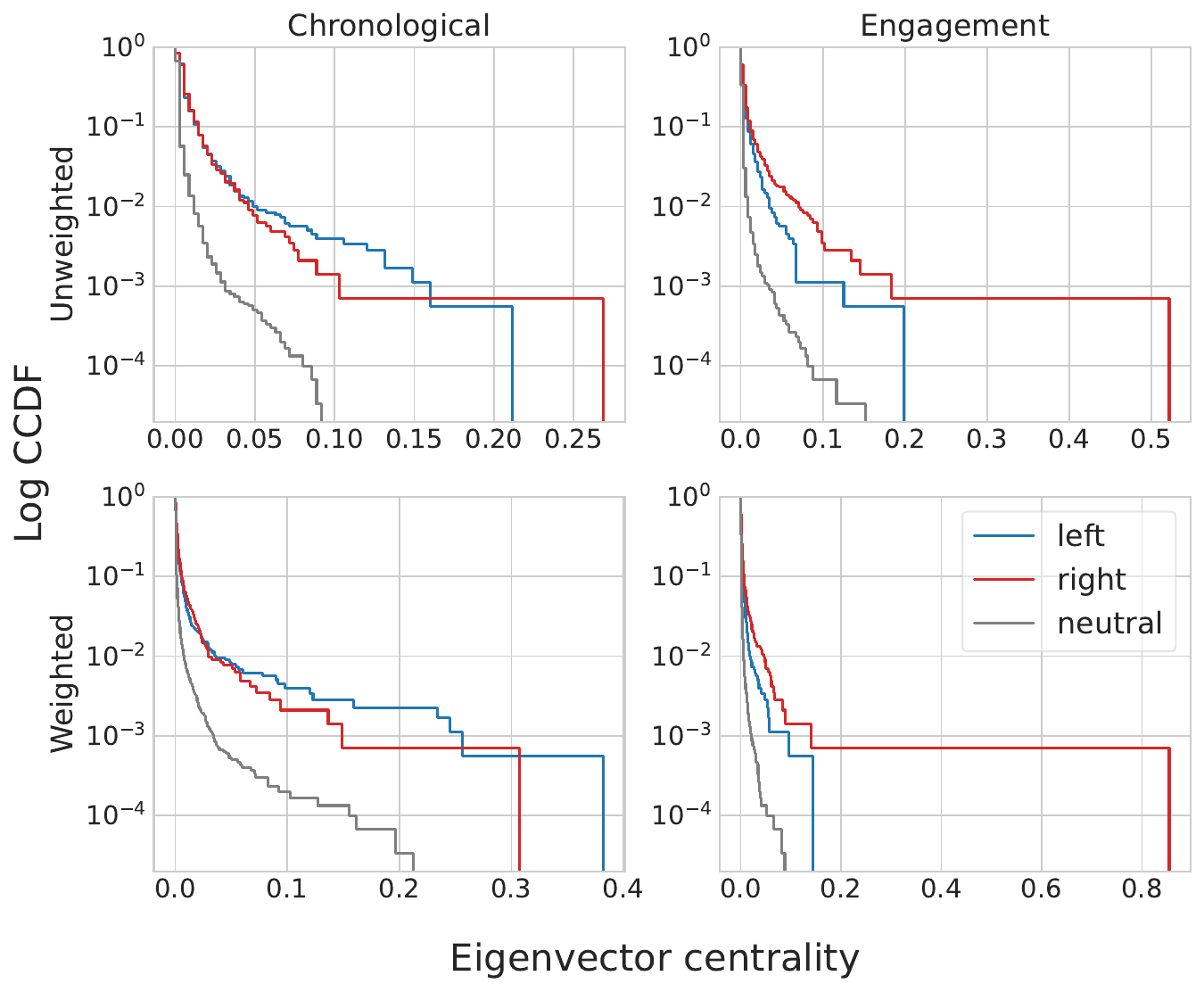}
        \caption{Eigenvector centrality CCDFs.}
        \label{subfig:ccdfs_eigen}
    \end{subfigure}
    \caption{CCDFs for node importance measures across different feeds.}
    \label{fig:ccdfs}
\end{figure*}

\paragraph{Top winners and losers.}
To better illustrate the kinds of accounts that featured most prominently in the chronological and engagement feed, we show the top-10 for each feed in Table~\ref{tab:feeds}.
This also demonstrates the ability of our method for determining political leaning to distinguish between well-known left- and right-leaning accounts on Twitter.
The table largely reflects the patterns in Figure~\ref{fig:ccdfs}; left-leaning accounts reached more users and were more important in the chronological feed, whereas the engagement feed featured right-leaning accounts more heavily.
We also notice a substantial difference in centrality distributions, with Elon Musk's account (the most influential in both feeds) becoming much more centralized in the engagement feed relative to the second most-central account (0.52 and 0.20, respectively) when compared to the chronological feed (0.27 and 0.21, respectively).
Thus, Elon Musk was likely the main driver of the substantially higher engagement feed centralization we report in Table~\ref{tab:net_stats}. 

\begin{table}[t!]
    \centering
    \small
    \begin{tabular}{lrr|lrr}
        \toprule
        \multicolumn{3}{c}{\textbf{Chronological}} & \multicolumn{3}{c}{\textbf{Engagement}} \\
        \midrule
        \textbf{Name} & \textbf{Deg.} & \textbf{C} & \textbf{Name} & \textbf{Deg.} & \textbf{C} \\
        \midrule
        \textcolor{red}{elonmusk} & 94 & 0.27 & \textcolor{red}{elonmusk} & 179 & 0.52 \\
        \textcolor{blue}{POTUS} & 74 & 0.21 & \textcolor{blue}{POTUS} & 68 & 0.20 \\
        \textcolor{blue}{nytimes} & 56 & 0.16 & \textcolor{red}{JackPosobiec} & 63 & 0.18 \\
        \textcolor{blue}{AP} & 52 & 0.15 & \textcolor{gray}{fasc1nate} & 52 & 0.15 \\
        \textcolor{blue}{TheOnion} & 46 & 0.13 & \textcolor{red}{hodgetwins} & 50 & 0.15 \\
        \textcolor{blue}{CNN} & 46 & 0.13 & \textcolor{red}{stillgray} & 46 & 0.13 \\
        \textcolor{blue}{washingtonpost} & 42 & 0.12 & \textcolor{blue}{RonFilipkowski} & 43 & 0.13 \\
        \textcolor{blue}{BBCWorld} & 37 & 0.11 & \textcolor{gray}{barstoolsports} & 40 & 0.12 \\
        \textcolor{red}{FoxNews} & 36 & 0.10 & \textcolor{red}{ClownWorld\_} & 35 & 0.10 \\
        \textcolor{blue}{JoeBiden} & 32 & 0.09 & \textcolor{red}{catturd2*} & 34 & 0.10 \\
        \bottomrule
    \end{tabular}
    \caption{Top-10 highest in-degree accounts in each (unweighted) feed. Colors indicate account leaning (red = right, blue = left, gray = neutral/unclassified). *Tied with DailyLoud, which is a right-leaning account.}
    \label{tab:feeds}
\end{table}

\begin{figure}[t!]
    \centering
    \includegraphics[width=0.99\linewidth]{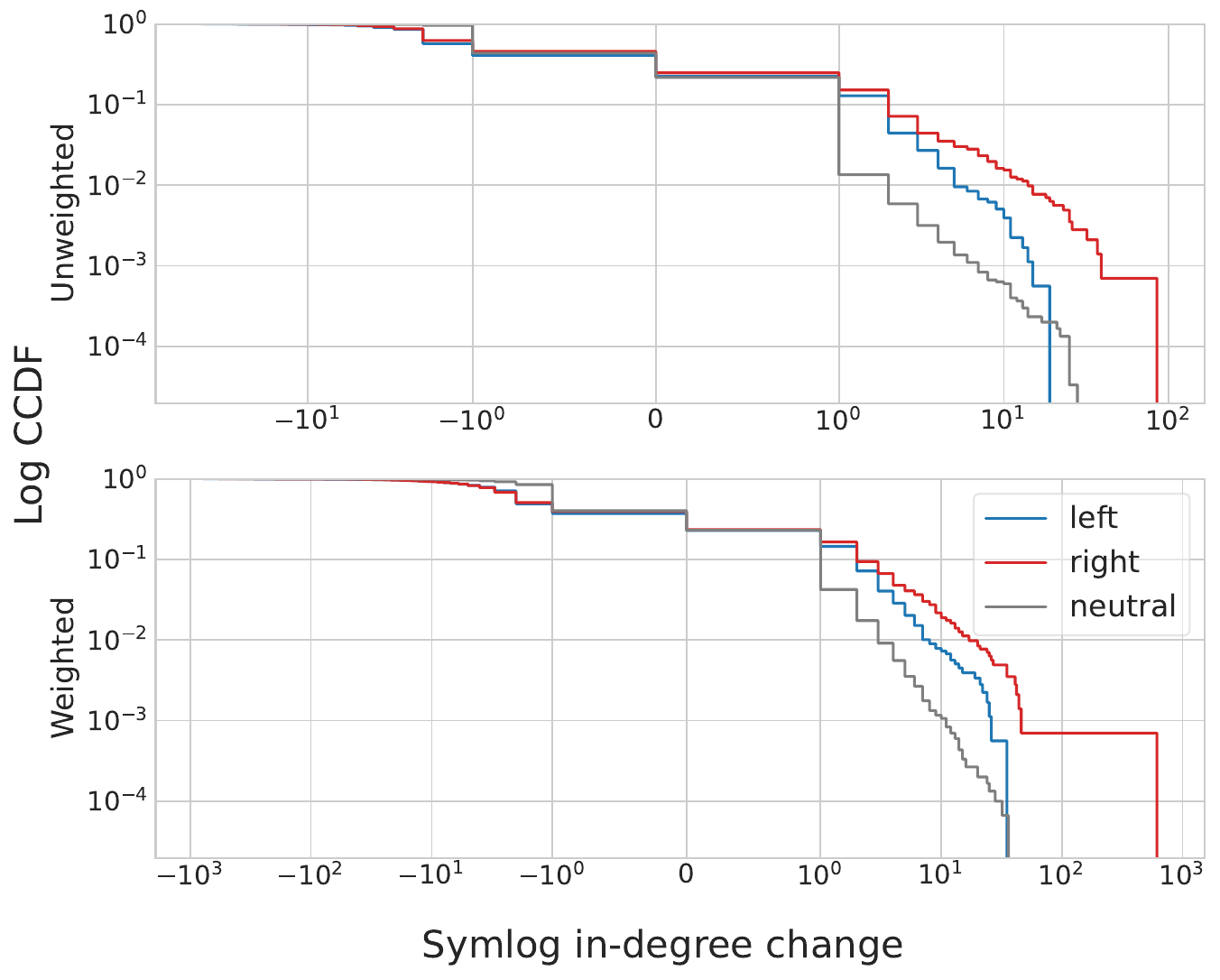}
    \caption{CCDF of degree change from chronological to engagement feed.}
    \label{fig:deg_gains}
\end{figure}

These right-leaning gains are also visible when we plot the relative in-degree change from the chronological to the engagement feed per account (Figure~\ref{fig:deg_gains}), where we also see some neutral/unclassified accounts gaining advantage over left-leaning accounts (especially in the unweighted network).
There are no discernible differences in terms of degree losses.
To further understand the nature of the accounts that gained and lost the most, we show the top 10 ``winners'' and ``losers'' in Table~\ref{tab:deg_deltas}; again, the patterns we observe are largely consistent in the non-balanced network.
We verify this rightward ``network seep'' in the Appendix, where we also consider whether accounts are in- or out-of-network (i.e., whether participants follow them or not); right-leaning accounts gained exposure and left-leaning accounts lost exposure in the engagement feed across self-identified Democrats, Republicans, and Independents alike.

\begin{table}[t!]
    \centering
    \small
    \begin{tabular}{lr|lr}
    \toprule
       \multicolumn{2}{c}{\textbf{Gains}} & \multicolumn{2}{c}{\textbf{Losses}} \\
       \midrule
        \textbf{Name} & $\Delta$ & \textbf{Name} & $\Delta$ \\
        \midrule
        \textcolor{red}{elonmusk} & +85 & \textcolor{blue}{TheOnion} & -43 \\
        \textcolor{red}{hodgetwins} & +39 & \textcolor{blue}{AP} & -40 \\
        \textcolor{red}{stillgray} & +37 & \textcolor{blue}{nytimes} & -33 \\
        \textcolor{red}{JackPosobiec} & +32 & \textcolor{blue}{BBCWorld} & -30 \\
        \textcolor{gray}{fasc1nate} & +28 & \textcolor{gray}{netflix} & -29 \\
        \textcolor{red}{DailyLoud} & +26 & \textcolor{blue}{washingtonpost} & -27 \\
        \textcolor{red}{bennyjohnson} & +25 & \textcolor{blue}{Reuters} & -27 \\
        \textcolor{red}{CollinRugg} & +25 & \textcolor{blue}{WhiteHouse} & -25 \\
        \textcolor{gray}{BornAKang} & +25 & \textcolor{blue}{NPR} & -23 \\
        \textcolor{gray}{HumansNoContext*} & +25 & \textcolor{blue}{CNN*} & -23 \\
        \bottomrule
    \end{tabular}
    \caption{Top-10 accounts with largest degree changes from (unweighted) chronological to engagement feed. *HumansNoContext and CNN were tied with vidsthatgohard (neutral) and SpaceX (right-leaning), respectively, but the tabulated accounts had more in-degrees in the engagement and chronological feeds, respectively.}
    \label{tab:deg_deltas}
\end{table}

From both Tables~\ref{tab:feeds} and~\ref{tab:deg_deltas}, one may derive that losses and gains in prominence from the chronological to the engagement feed were not purely a matter of political leaning.
Indeed, beneficiaries of the engagement feed seem to have been provocateurs or influencer-type accounts that may have posted more controversial content.
On the other hand, those that lost out were mostly news organizations (and other official accounts).
As such, despite clear gains for right-leaning accounts and losses for left-leaning ones, political leaning may be obfuscating the effects of other account characteristics like tweet tone. 
In the next section, we analyze account-level features and behaviors that may have been associated with increased prominence in the engagement feed.

\section{Algorithmically Rewarded Accounts}

To determine what kinds of accounts benefited the most from the engagement feed, we fit a regression model with several account-level characteristics (detailed below) as potential predictors and (unweighted) in-degree change as the outcome variable.

\paragraph{Qualifying $\Delta$in-degree as an outcome variable.}
We focus on in-degrees instead of eigenvector centrality because centrality values are dependent on the node's neighbors and the wider network topology, making centrality differences across feeds unintuitive.
Contrarily, in-degrees reflect the number of users that accounts were exposed to across feeds.
% or the number of account tweets that are shown overall for weighted and unweighted configurations, respectively.
We focus specifically on unweighted in-degrees, since weighted in-degrees may be more dependent on user preferences (e.g., users who followed fewer accounts would see those accounts more often in their chronological feeds) and risk being skewed by a few users.
% (e.g., the majority of weighted in-degrees for a single account could theoretically come from a single participant, which is problematic because our analyses are at the account, not the participant level).
Measuring in-degree \textit{change} allows us to normalize the power-distributed in-degrees across the two feeds (see next paragraph) which makes the regression coefficients meaningfully interpretable.
To control for the higher probability of a more-followed account appearing in a reverse-chronological feed (and thus potentially negatively influencing in-degree change), we add number of followers as a covariate in the model.
% However, this introduces another confound.
% If an account is followed by more people, we should reasonably expect that it has a higher probability of having a more negative in-degree change because it should appear in chronological feeds more often by default (as chronological feeds only show in-network accounts).
% For this reason, we add the number of followers as a predictor in our model to control for this possible negative effect.

\begin{figure}[t!]
    \centering
    \includegraphics[width=\columnwidth]{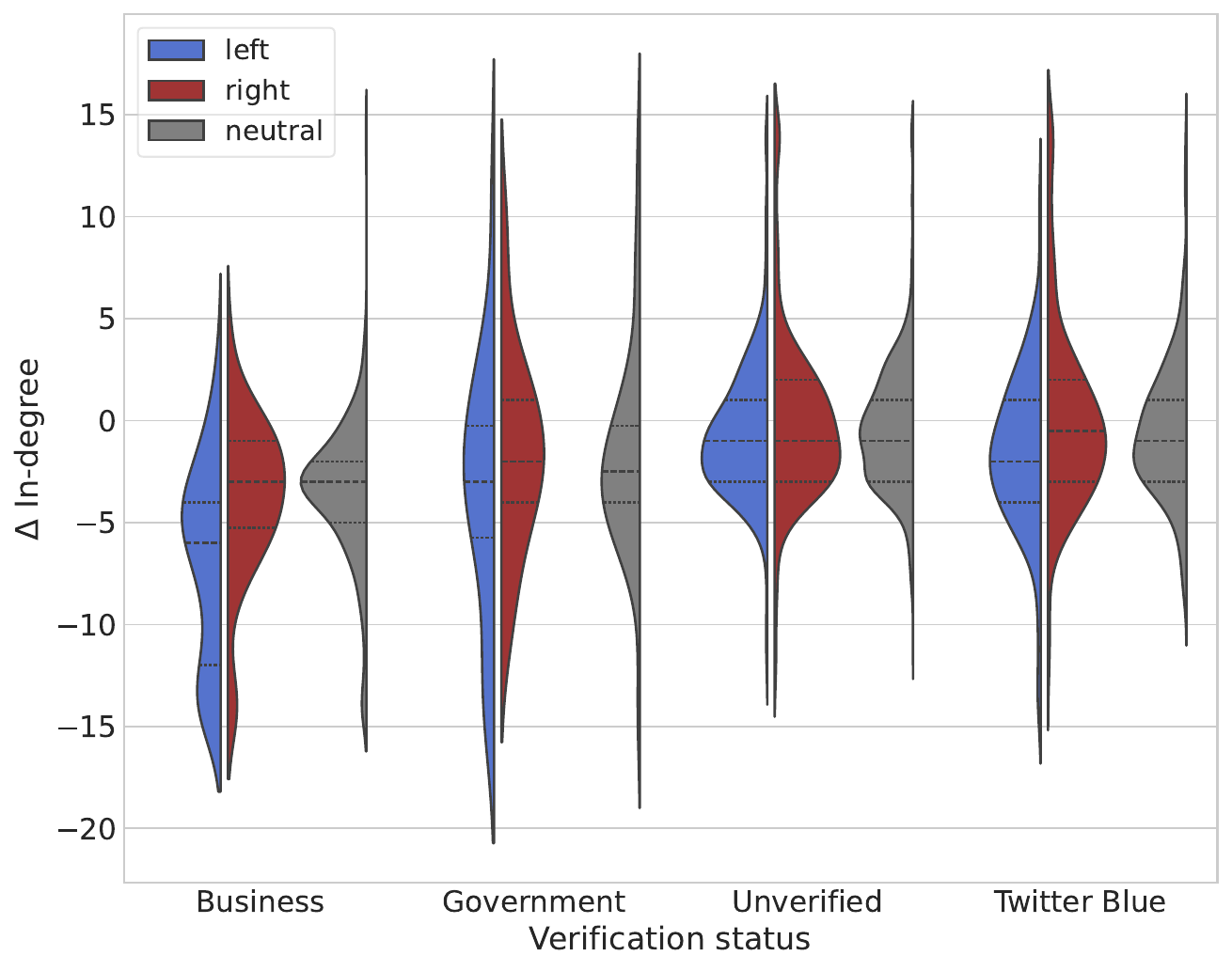}
    \caption{Violin plots by account political leaning and verification status.}
    \label{fig:violin}
\end{figure}

Before proceeding, we address the highly leptokurtic distribution of in-degree change by excluding any accounts followed by $< 3$ participants in the balanced network and winsorizing the 1\% most extreme values on either tail end.
% \alex{We can also do weighted degree analyses (raw exposures, not just users exposed to) -- but must account for participant effects (one participant can have disproportionate impact, this is also seen in much more skewed distributions for weighted in-degree changes)}
We show violin plots of the resulting distributions by account leaning and verification status in Figure~\ref{fig:violin}.
%(we outline how we determine verification status in the next subsection).
We select a minimum of 3 in-sample followers as a reasonable trade-off between minimizing the artificial inflation of neutral accounts and retaining an adequate portion of the account sample ($N$ = 2667).
Robustness checks with cutoffs of 4 and 5 result in identical effect directions.
Overall, as we increase the cutoff, we also observe increases in model fit ($R^2$) which increases confidence in our results.
However, there is also an inherent higher risk of overfitting due to a more limited sample.

\subsection{Account-Level Measurements}

We use a combination of existing account-level metrics, tweet-level metrics from the original dataset that we augment and transform into account-level variables, and other metrics that we derive based on our previous observations.

\paragraph{Account features.}
We consider the (log-transformed) number of overall followers that each account had, as well as their verification status.
To determine whether verification stemmed from Twitter Blue subscriptions or legacy verification, we use another dataset compiled with a combination of custom scraping and API queries.\footnote{\url{https://github.com/travisbrown/blue}}
The potential verification labels are no verification, business account, government account, or Twitter Blue subscriber.
We confirm that no accounts in the sample switched verification status during the data collection period.
In this feature category, we also consider the account's political leaning.

\begin{figure}
    \centering
    \includegraphics[width=0.99\linewidth]{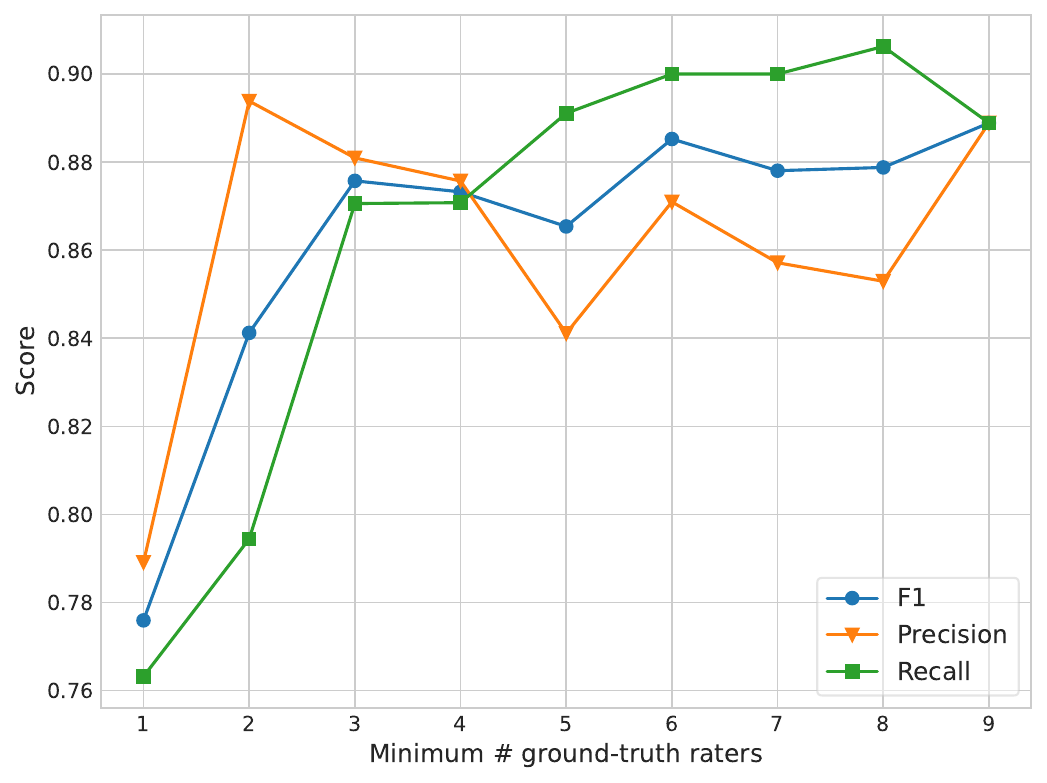}
    \caption{Gemini annotation performance by number of human annotators.}
    \label{fig:gemini}
\end{figure}

\paragraph{Tweeting style.}
For each tweet, we use an LLM (Gemini 2.0 Flash) to annotate whether it was political or not.
We test the model's performance against a subset of 30.6k tweets annotated by the participants in~\citet{milli_engagement_2025} (after removing URL-only tweets); the prompt we use for the LLM is adapted from the question shown to participants, which also captures social issues.
As we show in Figure~\ref{fig:gemini}, the model performs well and trends towards consensus; that is, performance is increased when considering tweets annotated by more participants that enable us to take the majority label.
For each account, we then compute the percentage of their in-sample tweets that were political.

We also use the same model to annotate whether a tweet was agitating/conflict-inducing (see Appendix for exact prompt).
% As with the political annotation prompt, we set model temperature to 0 and force structured output.
Although~\citet{milli_engagement_2025}'s dataset contains ratings for sentiment (e.g., anxiety, happiness) and anger expressed towards the left or right in a subset of tweets, analyses on GPT-4 labels provided in the original dataset reveal that LLM agreement with human annotators on these categories tends to be lower.
% A potential reason is that these categories are more context-dependent and may require more knowledge of the tweeting account's alignment or who the tweet is targeting (\textbf{NB:} The dataset responses are ordinal, not binary, but agreement continues to be low when binarizing them).
Thus, we opt for this custom \texttt{agitating} annotation, which we define as a tweet that stirs controversy or conflict.
We argue that this can capture more subtle ways of inducing negative responses and is more closely tied to the contents of the tweet itself, compared to commonly-used metrics like emotionality, toxicity, or identity attack which may also be largely dependent on the perceiver's identity~\cite{aroyo_crowdsourcing_2019,goyal_is_2022}.
For example, a tweet like ``Politician X is trying to destroy America'' could agitate both supporters and opponents of that politician alike; however, this tweet is neither toxic, nor does it directly attack any kind of identity.
% We opt for the \texttt{is\_agitating} label to follow up on our observations above as it relies only on information conveyed through the tweet itself.
% Whether the account uses agitation in a political context, and who the targets are, can then be inferred by the other account-level features we obtain.
% \alex{Qualify agitation metric here; better than toxicity because anger-inducing tweets may not necessarily be toxic, also more grounded in classic tactics like agitprop..}
% Moreover, we opt for agitation as opposed to other, more popular metrics like toxicity because we argue that this dimension can capture more subtle forms of 

As there is no agitation category in the original dataset, one of the authors annotates a random sample of 200 tweets while blinded to the model's labels to assess performance.
The LLM achieves an F1 score of 0.7 against this sample which is comparable to the single-rater analyses for the political category, indicating that it tends to correctly classify tweets.
As with the political labels, we compute each account's percentage of agitating in-sample tweets.
We note that both the political and agitating labels are based only on the tweet's \textit{text}, and not any other kinds of media (videos, images, etc.)

\paragraph{Proximity to network center.}
Given the high engagement feed centralization and the concentration of centrality around Elon Musk, another potential factor of visibility may have been proximity to Musk's account itself.
We operationalize this as whether Elon Musk interacted with a given account between his acquisition of Twitter on October 28th, 2022 and the end of the observation period.
We obtain all of Musk's tweets during this period and extract any accounts that he replied to or retweeted.
We do not consider quote tweets, because these are indistinguishable from original tweets in the dataset we use.
Since retweets only constituted $\sim$6\% of the remaining posts, we collapse both replies and retweets into a single interaction category.
We treat this as a binary variable of whether Musk interacted with the account in the given observation period or not.

\paragraph{External media.}
Several tweets contained external URLs, GIFs, photos, or videos.
%, either alongside text or without.
%; this information is provided by the original dataset.
To control for the potential effect of these, we also compute the average percentage of tweets containing each of them per account.

\subsection{Regression results}

\begin{figure}[t!]
    \centering
    \includegraphics[width=\columnwidth]{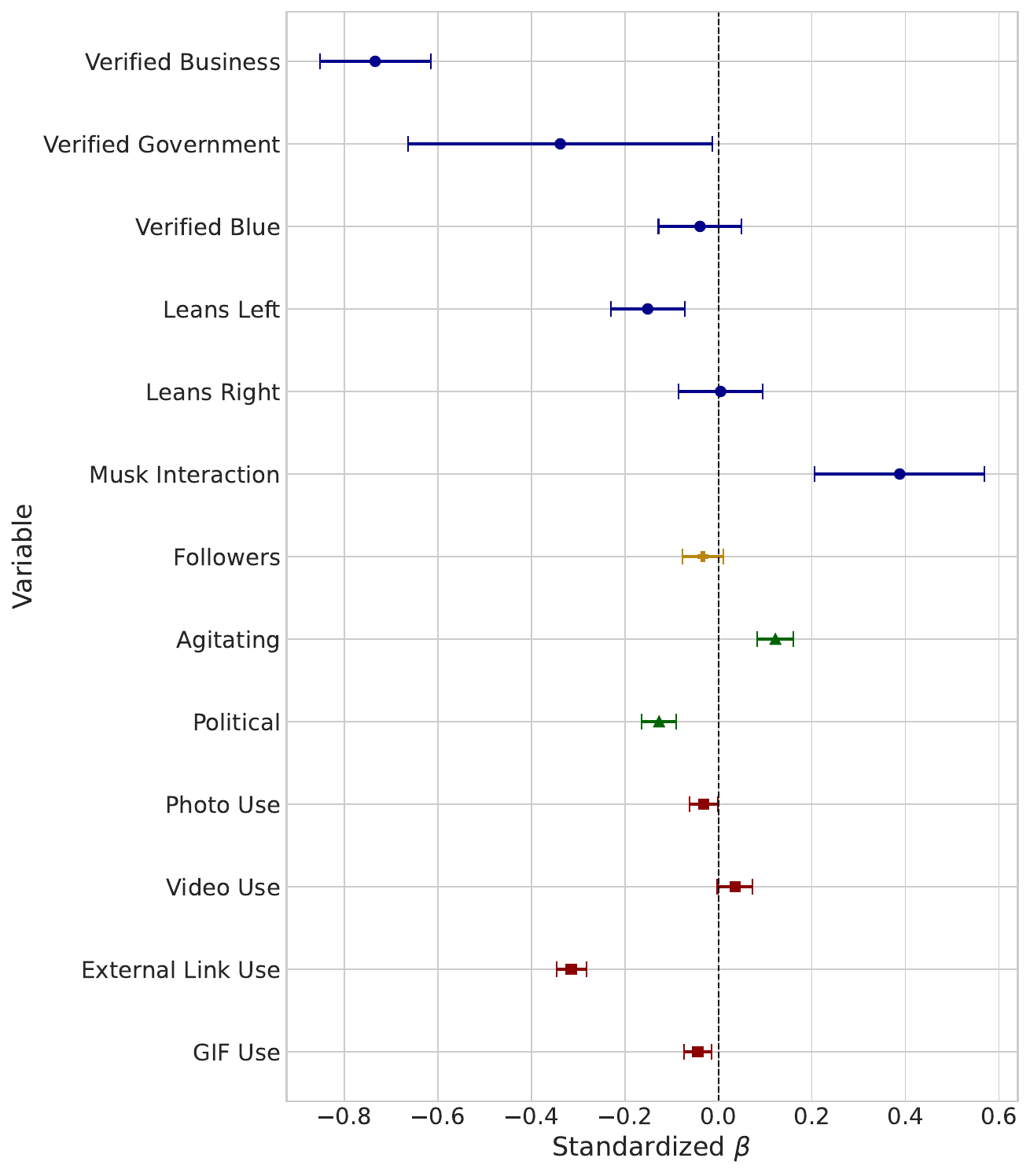}
    \caption{Standardized regression coefficients with 95\% CIs.}
    \label{fig:reg_coefs}
\end{figure}

We fit a multiple regression model with robust standard errors ($F_{(13,2653)} = 64.62, p < 0.001$, Adjusted $R^2 = 0.263$).
For the categorical variables account leaning, verification status, and Musk interaction, we use neutral, not verified, and no interaction as the reference categories, respectively.
In Figure~\ref{fig:reg_coefs}, we show the standardized $\beta$ coefficients with 95\% confidence intervals. 

Starting with the account-level features, we find that legacy verification, whether that was for an official Business ($p < 0.001$) or Government ($p = 0.04$) label, showed a significant loss of exposure in the algorithmically curated feed compared to being unverified.
Twitter Blue verification ($p = 0.39$) showed no effect.
A left political leaning resulted in a loss of prominence relative to neutral-leaning accounts ($p < 0.001$), whereas a right leaning showed no effect ($p = 0.92$).
The most positive influence was exerted by whether Elon Musk interacted with an account or not ($p < 0.001$), which corresponded to a large effect size (Cohen's $d = 0.93$).
In non-standardized terms, a Musk interaction corresponded to an average exposure of 1.5 more users in the algorithmic feed, which, based on our sample size of 376, translates to 3.99 new exposures per 1000 users.
We stress that this figure does not take into account any potential network effects (i.e., we assume a linear extrapolation).
We show non-standardized effects for all variables in the Appendix.

The (log) number of followers showed no effect ($p = 0.14$).
Regarding posting styles, we find that accounts posting more agitating tweets gained algorithmic exposure ($p < 0.001$) as opposed to accounts posting more political tweets, which lost exposure ($p < 0.001$).
% Alongside agitation, similarity to Elon Musk is the only other variable that positively predicts more algorithmic exposure ($p < 0.001$).

With the exception of videos, use of media in tweets was associated with loss in algorithmic exposure.
% ; however, this is not robust to using different cutoffs for in-sample followers for photo and GIF use.
Perhaps unsurprisingly, this loss was strongest for heavier use of external links, which may have been an artifact of the algorithm attempting to maximize user time spent on the platform ($p < 0.001$).

Overall, we confirm many of our previous observations.
Twitter's algorithm just prior to the platform's rebranding to X seems to have been rewarding accounts close to the platform's owner that tended to post conflict-stirring content.
On the contrary, it penalized official or other popular accounts, accounts that posted political content, and accounts that leaned left.
However, these per-variable effects are what we observe when the variance of all others is taken into account.
In the next section, we explore any potential interactions between some of these variables of interest.

\section{Differences Underlying Politics}

Although our results so far offer a characterization of what benefited an account in terms of algorithmic effects, there are many potential nuances and interactions that our regression model may not capture.
We reserve analysis of these interactions as post hoc data explorations to avoid an over-inflated model.
In this section, we provide descriptive accounts of some of the relationships between the independent variables themselves, as well as how they may have interacted to influence gains or losses in algorithmic exposure.
These analyses are aimed at providing a better understanding of what may have driven naive differences in the high-level gains of right-leaning accounts compared to left-leaning ones that we report above, and are not inferential.

\subsection{Musk Interactions by Leaning}

We begin with a simple cross-tabulation of interactions with Elon Musk by political leaning, shown in Table~\ref{tab:musk_ints} as observed vs.\ expected frequencies.
% In Table~\ref{tab:musk_ints}, we show the observed vs. expected frequencies of these interactions by leaning.
As can be seen, Elon Musk disproportionately interacted more with right-leaning accounts and less with left-leaning ones and (to a lesser extent) neutral ones.
These discrepancies are significant in a chi-squared test, $\chi^2 = 79.38, p < 0.001$.
It should be highlighted that accounts with which Elon Musk interacted were \textit{also} significantly more agitating compared to a randomly selected sample of equal size (\textit{N} = 217) that he did not interact with ($t_{(432)} = 4.23, p < 0.001$; statistical significance was robust with non-parametric tests).

\begin{table}[t!]
    \centering
    \small
    \begin{tabular}{lrrrr}
    \toprule
         & \multicolumn{2}{c}{\textbf{No interaction}} & \multicolumn{2}{c}{\textbf{Interaction}} \\
         \midrule
        \textbf{Acct. Lean} & \textit{E} & \textit{O} & \textit{E} & \textit{O} \\
        Left & 616.40 & 643 & 54.60 & 28 \\
        Neutral & 1293.44 & 1318 & 114.56 & 90 \\
        Right & 540.16 & 489 & 47.84 & 99 \\
        \bottomrule
    \end{tabular}
    \caption{Expected and observed frequencies of Musk interactions by leaning.}
    \label{tab:musk_ints}
\end{table}

We note that there are possible cascading effects resulting from this that we cannot capture here due to an incomplete network (randomly selected participants).
That is, if the most central account in the network interacted with mostly agitating, right-leaning accounts, that possibly increased their network centrality.
If these subsequently more central accounts interacted mostly with other right-leaning accounts, then those accounts are also likely to have benefited from neighboring important accounts, and so on.

\subsection{Agitation and Politicization by Leaning and Verification}

We perform two separate two-way (3x4) ANOVAs with robust standard errors using leaning, verification status, and an interaction term as independent variables to examine differences in the average political and agitating content that these accounts posted.
We provide descriptive statistics for these two high-level categories in Table~\ref{tab:lean_agpol}, and frequency cross-tabulations between them in Table~\ref{tab:xtabs}.
Since these tests are descriptive, we perform Type III ANOVAs where we continue to test for main effects even if an interaction is detected.

\begin{table}[t!]
    \centering
    \small
    \begin{tabular}{lrrrr}
    \toprule
        \textbf{Category} & $M_{pol}$ & $SD_{pol}$ & $M_{ag}$ & $SD_{ag}$ \\
        \midrule
        left & 0.49 & 0.39 & 0.22 & 0.28 \\
        neutral & 0.23 & 0.36 & 0.12 & 0.24 \\
        right & \textbf{0.51} & 0.39 & \textbf{0.30} & 0.30 \\
        \midrule
        business & 0.17 & 0.29 & 0.04 & 0.09 \\
        government & \textbf{0.77} & 0.31 & 0.21 & 0.25 \\
        unverified & 0.38 & 0.40 & \textbf{0.21} & 0.30 \\
        twitter blue & 0.36 & 0.40 & 0.20 & 0.26 \\
        \bottomrule
    \end{tabular}
    \caption{High-level descriptive statistics for political and agitating content by leaning and verification status.}
    \label{tab:lean_agpol}
\end{table}

\begin{table}[t!]
    \centering
    \small
    \begin{tabular}{lrrr}
    \textbf{Verification} & \textbf{Left} & \textbf{Neutral} & \textbf{Right} \\
    \toprule
       Business & 61 & 254 & 52 \\
       Government & 22 & 30 & 13 \\
       Not verified & 435 & 768 & 341 \\
       Twitter Blue & 153 & 356 & 182 \\
       \bottomrule
    \end{tabular}
    \caption{Cross-tabulation for category frequencies.}
    \label{tab:xtabs}
\end{table}

\paragraph{Agitation.}

\begin{figure*}[t!]
    \centering
    \begin{subfigure}[t]{0.49\textwidth}
        \centering
        \includegraphics[width=0.99\textwidth]{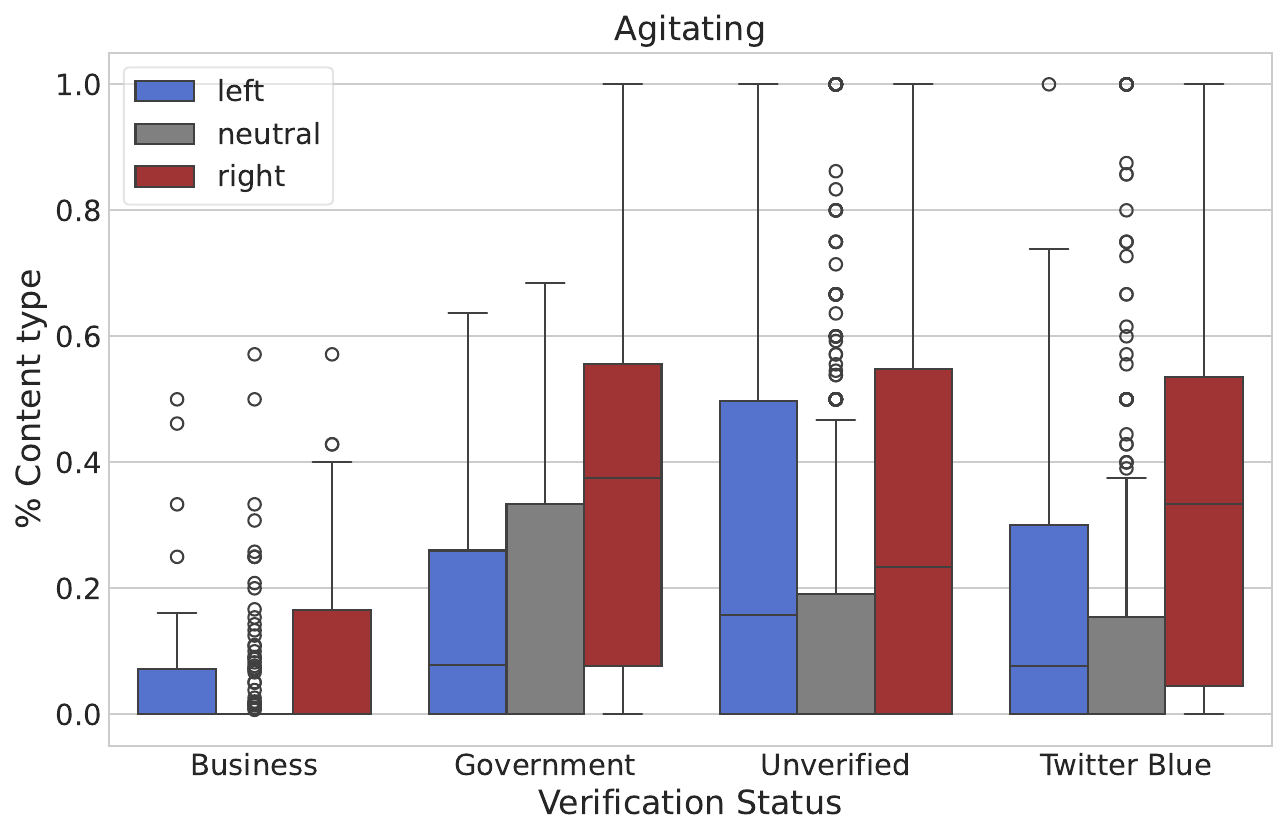}
        \caption{Agitating content.}
        \label{subfig:box_ag}
    \end{subfigure}
    \begin{subfigure}[t]{0.49\textwidth}
        \centering
        \includegraphics[width=0.99\textwidth]{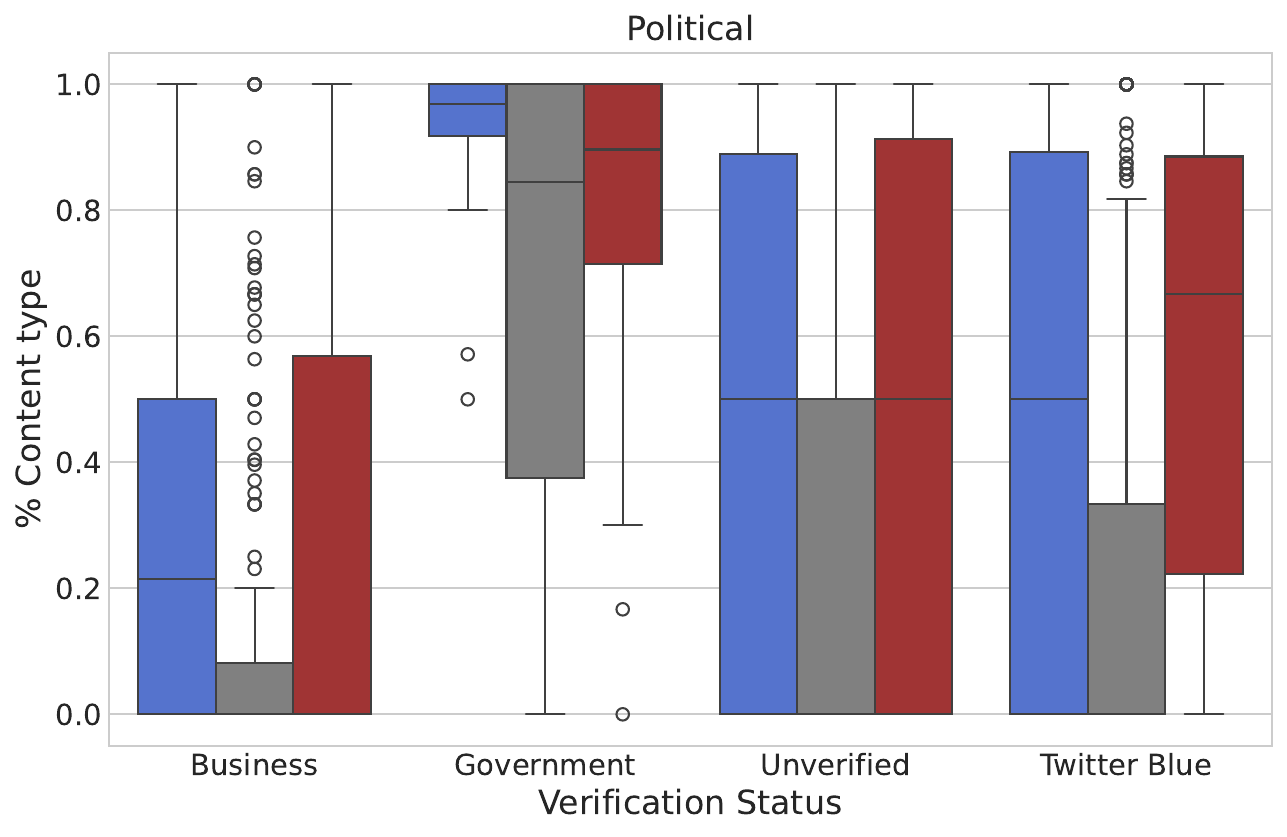}
        \caption{Political content.}
        \label{subfig:box_pol}
    \end{subfigure}
    \caption{Box plots by leaning and verification status breakdown.}
    \label{fig:box}
\end{figure*}

We find a significant interaction between leaning and verification status for agitation ($F_{(6,2655)}=6.18, p < 0.001$); Figure~\ref{subfig:box_ag} shows a descriptive breakdown of each group combination.
In Table~\ref{tab:tukey_ag}, we report results from pairwise Tukey's HSD tests to determine where these interactions lied by considering differences between political leanings for each level of verification status.

\begin{table}[t!]
    \centering
    \small
    \begin{tabular}{llrr}
    \toprule
        \textbf{Verification} & \textbf{Comparison} & $q$ & \textbf{95\% CI} \\
        \midrule
        \multirow{3}{*}{Business} & L-N & *0.034 & [0.004, 0.064] \\
         & L-R & -0.039 & [-0.079, 0.001] \\
         & N-R & ***-0.073 & [-0.105, -0.041] \\
         \midrule
         \multirow{3}{*}{Government} & L-N & 0.001 & [-0.165, 0.166] \\
         & L-R & -0.194 & [-0.399, 0.011] \\
         & N-R & *-0.195 & [-0.389, 0.000] \\
         \midrule
         \multirow{3}{*}{Not verified} & L-N & ***0.117 & [0.076, 0.158] \\
         & L-R & -0.046 & [-0.095, 0.003] \\
         & N-R & ***-0.163 & [0.118, 0.207] \\
         \midrule
         \multirow{3}{*}{Twitter Blue} & L-N & 0.040 & [-0.016, 0.097] \\
         & L-R & ***-0.170 & [-0.234, -0.106] \\
         & N-R & ***-0.210 & [-0.263, -0.157] \\
         \bottomrule
    \end{tabular}
    \caption{Post hoc Tukey HSD results for agitating content interactions. *$p < 0.05$, **$p < 0.01$, ***$p < 0.001$.}
    \label{tab:tukey_ag}
\end{table}

The post hoc comparisons reveal that, across all verification categories, right-leaning accounts always posted more agitating content than neutral ones.
Left-leaning accounts that were business-verified or unverified also posted more agitating content than corresponding neutral accounts.
In the case of Twitter Blue, right-leaning accounts posted more agitating content than left-leaning ones; we also observe a fairly substantial pattern for Government accounts in the same direction, although the small number of these Government accounts (see Table~\ref{tab:xtabs}) points to a likely lack of adequate statistical power to detect an effect.

These patterns are also observed in a significant main effect of political leaning ($F_{(2,2655)}=44.21, p < 0.001$).
Post hoc first-order Tukey HSD tests reveal that right-leaning accounts posted significantly more agitating content than both left ($q = 0.08, p < 0.001$) and neutral accounts ($q = 0.18, p < 0.001$); left-leaning accounts also posted more agitating content than neutral accounts ($q = 0.10, p < 0.001$).

For the significant main effect of verification status ($F_{(3,2655)}=62.11, p < 0.001$), post hoc analyses show that this was driven solely by business accounts that posted less agitating content than all three other categories (not verified, $q = -0.18, p < 0.001$; government, $q = -0.18, p < 0.001$; Twitter Blue, $q = -0.16, p < 0.001$).
None of the other pairwise comparisons showed significant differences.

\paragraph{Politicization.}

We repeat the same analyses for political content as the dependent variable (see Figure~\ref{subfig:box_pol} for descriptives).
Once again, we find a significant leaning-verification interaction ($F_{(2,2655)}=2.71, p = 0.013$), although this is more diminished compared to agitating content.
Table~\ref{tab:tukey_pol} shows the results from post hoc analyses.

\begin{table}[t!]
    \centering
    \small
    \begin{tabular}{llrr}
    \toprule
        \textbf{Verification} & \textbf{Comparison} & $q$ & \textbf{95\% CI} \\
        \midrule
        \multirow{3}{*}{Business} & L-N & ***0.182 & [0.088, 0.276] \\
         & L-R & 0.035 & [-0.089, 0.159] \\
         & N-R & **-0.147 & [-0.247, -0.047] \\
         \midrule
         \multirow{3}{*}{Government} & L-N & *0.238 & [0.039, 0.438] \\
         & L-R & 0.175 & [-0.073, 0.424] \\
         & N-R & -0.063 & [-0.299, 0.173] \\
         \midrule
         \multirow{3}{*}{Not verified} & L-N & ***0.247 & [0.193, 0.301] \\
         & L-R & -0.009 & [-0.074, 0.057] \\
         & N-R & ***-0.256 & [-0.314, -0.197] \\
         \midrule
         \multirow{3}{*}{Twitter Blue} & L-N & ***0.263 & [0.179, 0.346] \\
         & L-R & -0.086 & [-0.181, 0.009] \\
         & N-R & ***-0.349 & [-0.427, -0.270] \\
         \bottomrule
    \end{tabular}
    \caption{Post hoc Tukey HSD results for political content interactions. *$p < 0.05$, **$p < 0.01$, ***$p < 0.001$.}
    \label{tab:tukey_pol}
\end{table}

As expected, left-leaning accounts posted more political content than neutral ones across all verification categories; right-leaning accounts also posted more political content than neutral ones for all verification types except for government.
We observe no significant differences in prominence of political content between left and right in any of the verification categories.
As with previous analyses, we may lack statistical power for the government category, where left-leaning accounts had a higher mean than right-leaning ones.  

These findings are also reflected in the significant main effect of political leaning ($F_{(2, 2655)} = 82.36, p < 0.001$).
Post hoc tests reveal significant differences between left and neutral ($q = 0.26, p < 0.001$) and right and neutral ($q = 0.28, p < 0.001$), but not between right and left ($q = 0.019, p = 0.65$).

In addition, there is a significant main effect of verification status ($F_{(3, 2655)} = 31.25, p < 0.001$).
Naturally, government accounts posted substantially more political content than all other categories (not verified, $q = 0.399, p < 0.001$; business, $q = 0.605, p < 0.001$; Twitter Blue, $q = 0.412, p < 0.001$).
Both not verified ($q = 0.205, p < 0.001$) and Twitter Blue ($q = 0.192, p < 0.001$) accounts posted more political content than business accounts.
We see no substantial differences between not verified and Twitter Blue ($q=0.013, p = 0.881$).

Overall, right-leaning accounts tended to post more agitating content than neutral or left-leaning accounts, particularly when they were Twitter Blue-verified. 
Left-leaning accounts, and especially non-verified ones, also posted more agitating content than neutral accounts.
For verification status, effects were driven by business accounts that tended to post less agitating content than other types, perhaps as a form of brand safety.

As expected, left- and right-leaning accounts posted more political content than neutral ones regardless of verification type (with the exception of right-neutral comparisons for government accounts); between left and right, politicization levels tended to be similar.
Notably, posting more political content also correlated with posting more agitating content ($\rho = 0.61, p < 0.001$).

For additional context, we also show a three-way analysis of the interaction between agitation, politicization, and political leaning and its effects on visibility changes between feeds in the Appendix.

% \subsection{Musk Similarity by Leaning}
% \alex{change subsection heading to musk interaction by leaning}

% \alex{probably best to move this paragraph somewhere else, also should maybe do this by verification status too (esp. interesting to look at Twitter Blue but depends on how we operationalize Clemson data)}

% Given the importance of the MSI metric in the overall regression model and the high-level political differences in gain and loss of exposure, we also perform a one-way ANOVA with political leaning as the independent variable to determine whether there are differences in account similarity to Elon Musk.
% We find a significant main effect of political leaning on MSI, $F_{(2, 2665)}=89.65, p < 0.001$.
% Following pairwise Tukey HSD post hoc tests, we find that this difference is driven solely by a higher similarity of right-leaning accounts compared with both left-leaning ($q = 0.195, p <0.001$) and neutral ($q = 0.211, p < 0.001$) accounts.
% We observe no significant differences between left and neutral ($q = 0.016, p = 0.555$).

\section{Discussion}

Our findings shed light on how Twitter's recommendation algorithm prior to its re-branding to X may have affected the visibility of different accounts.
While we replicate the finding reported by several works~\cite{graham_computational_2024,huszar_algorithmic_2022,Duskin2026-new, ye_auditing_2025} that right-leaning accounts benefited from algorithmic curation, upon closer inspection, we find that this was likely due to them acting in ways that correlated with algorithmic rewards.
Namely, right-leaning accounts posted more agitating content and were closer to Elon Musk, both of which were associated with more algorithmic exposure.

At the same time, contrary to previous work~\cite{huszar_algorithmic_2022}, we find that accounts of government officials (and businesses, which are also legacy-verified) lost visibility in the algorithmic configuration.
There could be several reasons behind this, including different timelines, operationalization of government accounts, and the fact that~\citet{huszar_algorithmic_2022} did not only focus on the US context.
% the fact that we analyze data collected after Twitter's algorithm underwent substantial changes \alex{check Huszar timeline}, and differences in how \citet{huszar_algorithmic_2022} operationalized government official accounts \alex{check the measurement}.

\paragraph{Implications.}
Among the biggest strengths of our work is the examination of algorithmic account visibility during a particularly pivotal time for the platform.
The data we analyze were collected right after a change in Twitter's leadership, but right before its re-branding to $\mathbb{X}$.
Moreover,~\citet{milli_engagement_2025}'s collection coincided with the release of Twitter's algorithm, which enables a direct comparison between the algorithm's stated purpose and realized effects.
For example, the algorithm's code contained identifying flags for whether a user was a Democrat or Republican, or even for whether a tweet was authored by Elon Musk.\footnote{\url{https://archive.ph/KdGqX}}
Though these flags were likely used for testing and monitoring, they echo our findings in how account characteristics carry important implications for content visibility.
We perform account-level analyses to uncover the intricacies of such algorithmic design choices, in contrast with prior work focusing on content-level characteristics \cite{milli_engagement_2025}. 

Additionally, our findings highlight a troublesome implication.
In the pursuit of reaching wider audiences, users may be incentivized to stir controversy or vie for engagement with the platform's owner, creating an environment of elevated agitation and inadvertent ``permissible reach''.
Indeed, some work has demonstrated an uptick in problematic behavior such as hate speech and automated activity following Musk's acquisition of the platform~\cite{hickey_auditing_2023,hickey_x_2025}.
This is antithetical to the platform's ostensible role as a ``digital town square'', especially in the wake of cuts to the Trust and Safety team, which would perhaps be best placed to monitor and address such issues~\cite{moran_end_2025_new}.

Moreover, we add to literature demonstrating political differences in the adoption of problematic behaviors~\cite{mosleh_differences_2024}, which possibly explains the different rates at which different political groups are subjected to content moderation~\cite{haimson_disproportionate_2021,renault_republicans_2025}.
Thus, our work helps to further contextualize research on the types of content that achieve more visibility~\cite{galeazzi_revealing_2026}, as well as discussions around why accusations of bias may ignore crucial context behind these moderation decisions.\footnote{\url{https://www.techpolicy.press/scientists-respond-to-ftc-inquiry-into-tech-censorship/}}

\paragraph{Limitations.}
Although we perform comparisons between feeds for the same users, our findings are not necessarily causal.
Without systematic randomization of users into different configurations~\cite{guess_how_2023,huszar_algorithmic_2022} or counterfactual behaviors~\cite{hosseinmardi_causally_2024} it is difficult to disentangle algorithmic amplification from overall user preferences.
However, given that we are essentially comparing ``counterfactual feeds'', these findings are a good description of how user experience differed in algorithmic versus reverse-chronological configurations.

Moreover, due to the short and specific timespan of the data we analyze, our results may not generalize beyond this period.
We cannot confidently state that the effects we report would extend to pre-Musk or post-$\mathbb{X}$ times, though we report several similarities with works from those periods~\cite{huszar_algorithmic_2022,ye_auditing_2025}.
Nonetheless, we reiterate the intrigue of this particular period as it is close to the date on which Twitter released its recommendation algorithm, allowing for more insight into the algorithm's stated versus realized behavior.

\section{Acknowledgments}

This work has been supported by the University of Washington’s Center for an Informed Public, the John S. and James L. Knight Foundation (G-2019-58788), and the William and Flora Hewlett Foundation (2023-02789).
The authors thank Akhil Chennamsetty and Darren Linvill for providing the Musk interaction data.

% \small
\bibliography{cleaned_references,references_betb}
% moved new refs to old bib file
% \bibliography{extras.bib}

\normalsize
\clearpage

\clearpage
\appendix
\section{Appendix}

\subsection{Verification of Matching Procedure}\label{app:matching}

We assess our matching procedure by verifying that participant demographic distributions are closer post-matching compared to pre-matching.
Since we are working with ordinal and categorical data, and therefore cannot implement standardized mean differences, we instead compute the Jensen-Shannon divergence between each variable's distribution from left-leaning and right-leaning users.
We show the results for the whole left-leaning and matched left-leaning sample in Table~\ref{tab:app_jsd}, which demonstrates that we obtain drastically higher distribution equality (lower $D_{JS}$ for gender and modestly higher distribution equality for race, reason for using Twitter, and age without sacrificing substantial equality on education (which is slightly more unequal in the matched pairs) and income (which is approximately equal between matched and unmatched).

\begin{table}[h]
    \centering
    \small
    \begin{tabular}{lrr}
    \toprule
         \textbf{Variable} & $D_{JS}(L_{pre}||R)$ & $D_{JS}(L_{post}||R)$ \\
         \midrule
         Race & 0.155 & \textbf{0.101} \\
         Gender & 0.134 & \textbf{0.043} \\
         Education & \textbf{0.083} & 0.092 \\
         Twitter Reason & 0.077 & \textbf{0.031} \\
         Age & 0.087 & \textbf{0.030} \\
         Income & 0.073 & \textbf{0.071} \\
         \bottomrule
    \end{tabular}
    \caption{Jensen-Shannon divergences between left- and right-leaning demographic distributions for pre- and post-matching participants.}
    \label{tab:app_jsd}
\end{table}

\subsection{Activity Over Time}

\begin{figure*}[t!]
    \centering
    \includegraphics[width=0.95\linewidth]{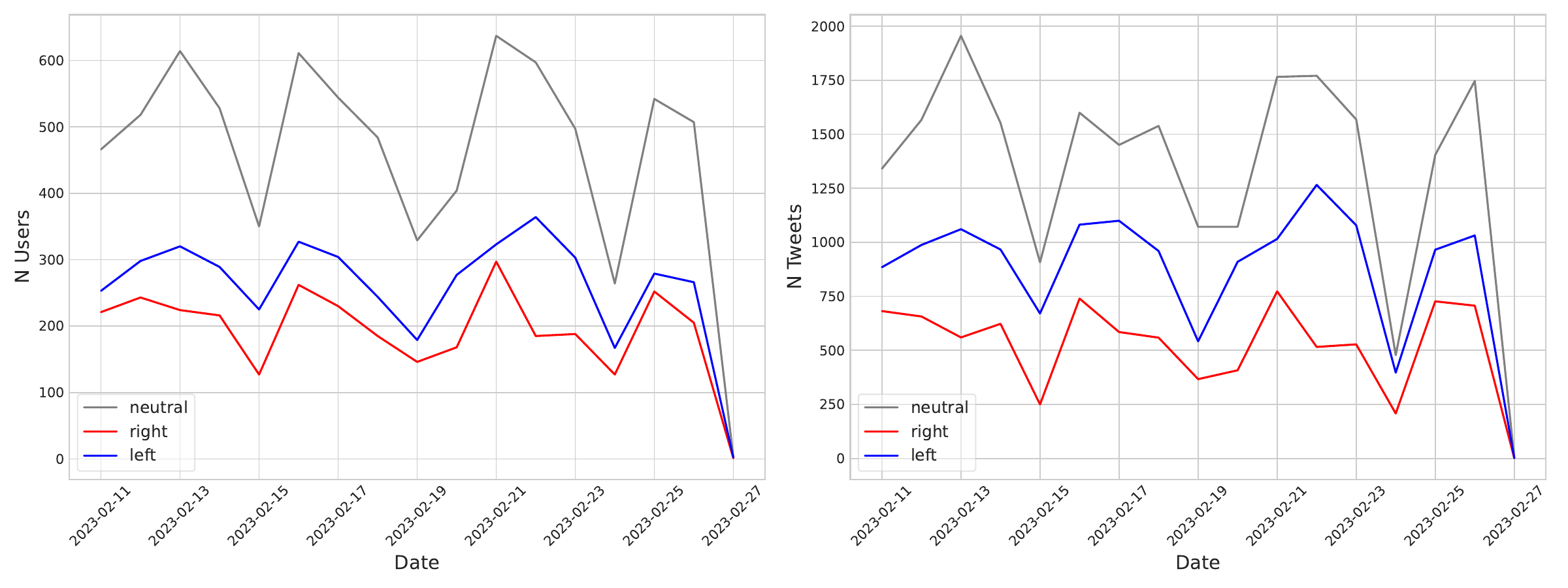}
    \caption{Daily activity (number of unique users and tweets) by political leaning.}
    \label{fig:app_tsactivity}
\end{figure*}

We show time-series plots of the volume of unique accounts and tweets posted across the February 2023 observation period by political leaning in Figure~\ref{fig:app_tsactivity}.
Activity patterns are largely similar across left, right, and neutral accounts, with no unexpected drop-offs for any particular leaning.

\subsection{Network Seeping}

The reverse-chronological feed is distinguished from the engagement feed by virtue of it featuring only in-network accounts (i.e., accounts participants follow) as opposed to the engagement feed that also features out-of-network ones.
We show the cross-tabulation of the number of accounts that appear in the different feeds by whether the participant to whom the feed belongs follows them or not in Table~\ref{tab:app_xtabnet}.
This confirms that in-network accounts are much more prominent in the chronological feed.
Note that the out-of-network accounts in the chronological feed are due to the way the data are logged; retweets by an in-network account still show the (potentially out-of-network) original tweet author as the account that posted the tweet.

\begin{table}[t!]
    \centering
    \small
    \begin{tabular}{lrr}
    \toprule
        \textbf{Feed} & \textbf{Not following} & \textbf{Following} \\
        \midrule
        Chronological & 4.8k & 7.4k \\
        Engagement & 3.9k & 3.4k \\
        \bottomrule
    \end{tabular}
    \caption{In- and out-of-network accounts per feed.}
    \label{tab:app_xtabnet}
\end{table}

To fix the patterns we report in the main paper to whether accounts appearing in feeds are in- or out-of-network, we set the expected number of accounts that participants will see per leaning on the proportion of account leanings that they follow.
We then observe the actual number of account leanings that appear in their engagement feeds, and determine the deviation between these frequencies using chi-squared tests.
For this analysis, we look at participants' parties rather than their political leaning, as Independents may be a special interest category in this case.
We run these tests on the entire participant sample, not just the matched one, as chi-squared makes no balance assumptions.
We show the (follow-based) expected and (engagement feed-based) observed frequencies in Table~\ref{tab:app_freqs}.
Note that we rescale the observed frequencies so that row-wise sums match the expected frequencies.

\begin{table}[t!]
    \centering
    \small
    \begin{tabular}{lrrrrrr}
    \toprule
         & \multicolumn{2}{c}{\textbf{Democrat}} & \multicolumn{2}{c}{\textbf{Independent}} & \multicolumn{2}{c}{\textbf{Republican}} \\
         \midrule
        \textbf{Acct. Lean} & \textit{E} & \textit{O} & \textit{E} & \textit{O} & \textit{E} & \textit{O} \\
        Left & 2.75k & 2.38k & 851 & 683 & 120 & 86.7 \\
        Neutral & 2.86k & 2.79k & 1.37k & 1.34k & 813 & 728 \\
        Right & 420 & 860 & 625 & 819 & 941 & 1.06k \\
        \bottomrule
    \end{tabular}
    \caption{Expected frequencies based on number of accounts following per leaning and (scaled) observed frequencies of appearances in the engagement feed.}
    \label{tab:app_freqs}
\end{table}

We see that, consistently, left-leaning accounts feature less frequently than would be expected based on followers in the engagement feed, while right-leaning accounts feature more frequently; neutral accounts also feature slightly less frequently.
As we show in Table~\ref{tab:app_chi2}, these discrepancies are statistically significant in chi-squared tests for all three categories of participant party, with the most substantial discrepancy being a heavy featuring of right-leaning accounts in Democrat feeds.
Therefore, we confirm our observations in the main paper as not being merely due to discrepancies in baseline proportions of the account leanings that different participants follow.

\begin{table}[t!]
    \centering
    \small
    \begin{tabular}{lrrr|r}
    \toprule
         & \multicolumn{3}{c}{\textbf{Account leaning} $\%_{diff}$} &  \\
         \midrule
        \textbf{Party} & \textbf{Left} & \textbf{Neutral} & \textbf{Right} & $\chi^2$ \\
        Democrat & \textcolor{red}{-13.58\%} & \textcolor{red}{-2.34\%} & \textcolor{teal}{+104.9\%} & *514.24 \\
        Independent & \textcolor{red}{-19.76\%} & \textcolor{red}{-1.92\%} & \textcolor{teal}{+31.11\%} & *94.21 \\
        Republican & \textcolor{red}{-27.73\%} & \textcolor{red}{-10.50\%} & \textcolor{teal}{+12.61\%} & *33.14 \\
        \bottomrule
    \end{tabular}
    \caption{Percentage differences and chi-squared statistics between follow-expected and feed-observed frequencies. Party refers to participants' self-reports. *$p<0.001$ (all significant at this level).}
    \label{tab:app_chi2}
\end{table}

\subsection{Gemini Agitation Prompt}

For agitation labels, we use the following prompt:

\begin{quote}
    You are a research assistant. For each subsequent text you receive, you must answer this question: Is this tweet stirring up conflict? Return your answer in JSON format with key ``is\_agitating'' and value either ``yes'' or ``no''.
\end{quote}

\subsection{Non-standardized Regression Effects}

\begin{figure}[t!]
    \centering
    \includegraphics[width=0.85\columnwidth]{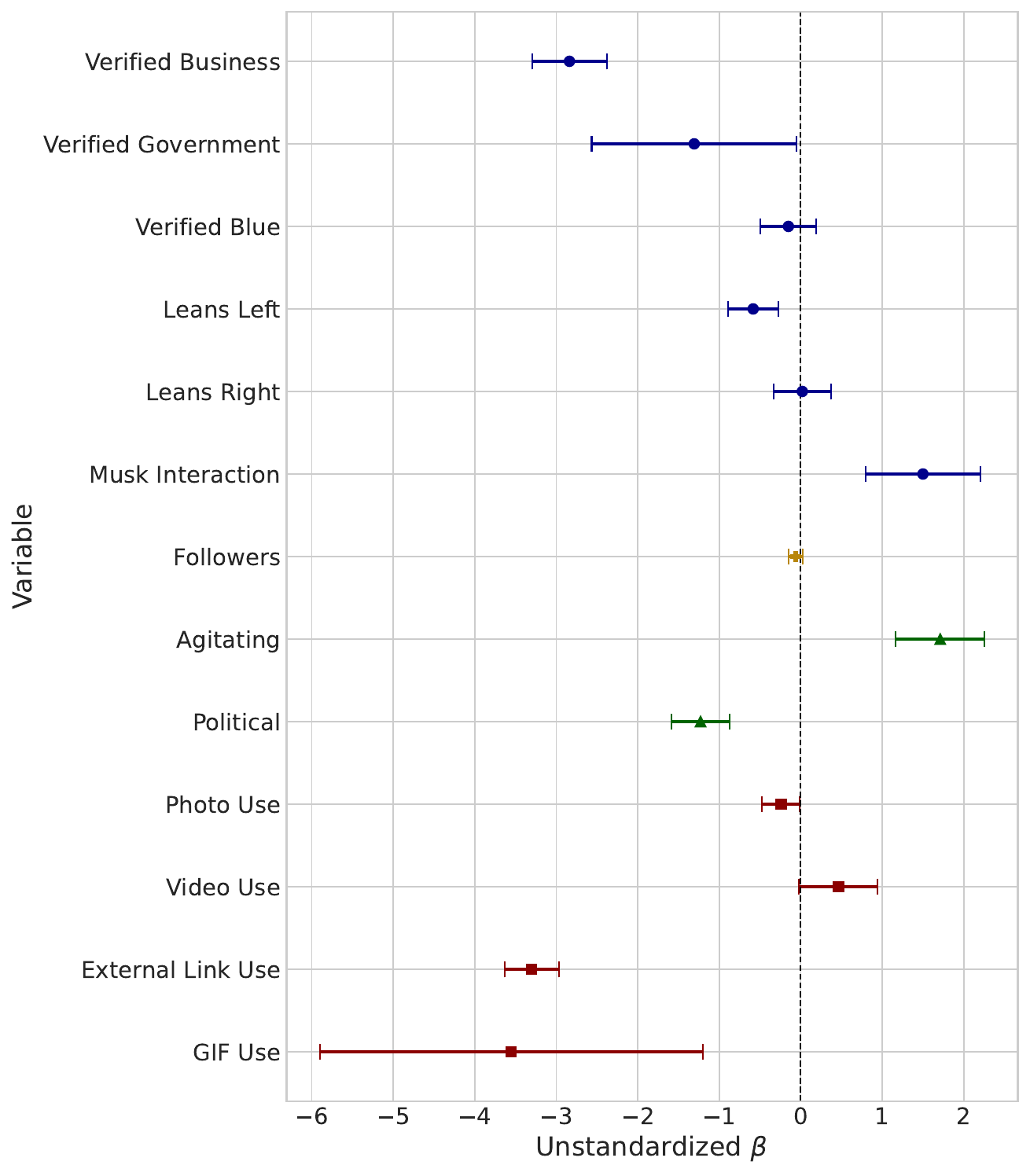}
    \caption{Non-standardized coefficients with 95\% CIs.}
    \label{fig:reg_nonstd}
\end{figure}

In Figure~\ref{fig:reg_nonstd}, we show the non-standardized coefficients of each regressor for easier interpretation of marginal effects, as opposed to the relative effects shown in Figure~\ref{fig:reg_coefs}.

\subsection{Interactions on Algorithmic Exposure}

In this analysis, we focus on how the three-way relationship between agitation, politicization, and political leaning, is associated with visibility changes between feeds.
To make our interpretation of any potential relationships intuitive, we binarize the agitation and politicization variables by splitting accounts on the median.
Note that the median for agitation is 0, therefore, any account with at least one agitating tweet in the observation period is treated as agitating for the purposes of this exploratory analysis.

\begin{figure}[t!]
    \centering
    \includegraphics[width=0.9\linewidth]{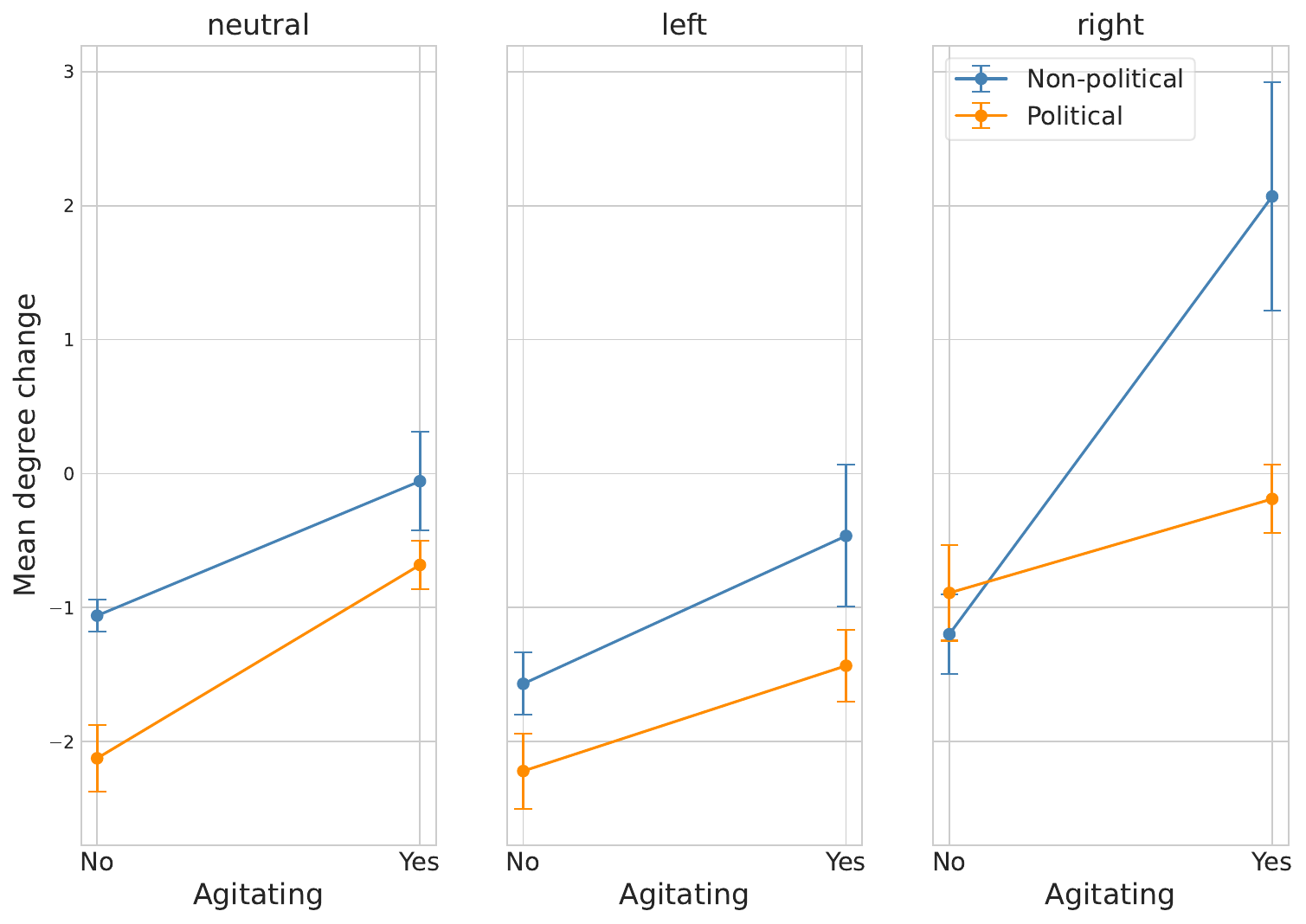}
    \caption{Interaction visualizations between account politicization, agitation, and leaning. Note that degree change always trends negative due to fewer overall posts in the engagement feed as opposed to the reverse-chronological feed.}
    \label{fig:agpol_int}
\end{figure}

When fitting a 3x2x2 ANOVA with these transformations, we find a significant 3-way interaction ($F_{(2,2655} = 3.54, p=0.03$) which indicates differential effects across the different levels.
We plot this interaction in Figure~\ref{fig:agpol_int}, which follows the general effects we find in the full regression model; political accounts show lower degree gains (or higher losses), and agitating ones have higher (or less negative) degree gains.

However, there are also interesting patterns in the interactions.
For neutral accounts, the gap between increased exposure for non-political versus political accounts is narrowed when their content is agitating.
For right-leaning accounts, we observe no difference in exposure for political and non-political accounts when their content is not agitating; however, there is an uptick for exposure of accounts that post agitating but non-political content, above and beyond all other possible category combinations.

\end{document}